\documentclass[11pt,colorlinks=true,citecolor=blue,urlcolor=blue,linkcolor=blue]{article}
\usepackage[letterpaper,textwidth=6.5in,textheight=9.0in]{geometry}

\usepackage[numbers,sort&compress]{natbib}
\usepackage{graphicx}
\usepackage{amsmath,amssymb,bm}
\usepackage{booktabs}
\usepackage{physics}
\usepackage{array}
\usepackage{placeins}
\usepackage{url}
\usepackage{doi}
\usepackage{hyperref}

\renewenvironment{abstract}{\quotation}{\endquotation}

\date{}

\def\scititle{Single-photon time qubit: Quantum-vacuum control and complementarity of the arrow of time}

\title{\bfseries \boldmath \scititle}

\author{%
Tai Hyun Yoon$^{1,2*}$\\
\small $^{1}$Department of Physics, Korea University, Seoul 02841, Republic of Korea\\
\small $^{2}$Center for Molecular Spectroscopy and Dynamics, Institute for Basic Science,\\
\small Korea University, Seoul 02841, Republic of Korea\\
\small $^{*}$Corresponding author. Email: thyoon@korea.ac.kr
}

\begin{document}
\maketitle

\begin{abstract}\bfseries\boldmath
Treating the arrow of time as a controllable quantum variable rather than a fixed
external parameter is a central challenge in quantum physics. Recently,
Bloch~\cite{Bloch2025} proposed the theoretical concept of a \emph{time qubit}, a
two-level control space $\mathcal{H}_T=\mathrm{span}\{\ket{+},\ket{-}\}$ that
selects between opposite Hamiltonian evolutions $\hat{H}_{\pm}=\hat{H}_0\pm\hat{V}$,
but this paradigm has remained experimentally unrealized. Here we formulate the
quantum-optical theory of a single-photon time qubit---uncovering quantum-vacuum
control and complementarity of the arrow of time---and experimentally probe and
validate its predictions using phase-resolved single-photon interferometry from
two coherently seeded entangled nonlinear biphoton sources~\cite{Lee2018,Yoon2021}.
First, we show theoretically and verify experimentally that a controlled $\pi$ pump--seed phase inversion during pair creation from the quantum vacuum flips the stimulated displacement generator $\hat{V}\to -\hat{V}$ on top of the spontaneous two-mode squeezing background $\hat{H}_0$, reversing the stimulated evolution time [$\hat{U}_{\mathrm{stim},-}(t)=\hat{U}_{\mathrm{stim},+}(-t)$] and enabling universal single-qubit $\mathrm{SU}(2)$ control and postselection-free
first-order tomography on the temporal Bloch sphere. Second, we prove that spontaneous idler vacuum fluctuations record complete which-branch (temporal-direction) information whereas the shared coherent seed erases it without cascading or aligning the idler beams, establishing an exact
wave--particle complementarity law $|\bm{r}|=\mathcal{V}=|\alpha|^2/(1+|\alpha|^2)$
for the purity of the arrow of time and enabling direct first-order measurement
of the Loschmidt time-reversal echo.
\end{abstract}

\section*{Introduction}

Time enters standard quantum mechanics in a way that differs sharply from other
physical observables. While position, spin, and field quadratures are represented
by self-adjoint operators that can be measured projectively, time is typically
introduced as an external $c$-number parameter labeling unitary evolution, a
structural asymmetry rooted in Pauli's theorem~\cite{Pauli1933} (see
Supplementary Materials, Section~\ref{sec:S2_Pauli}). This asymmetry has
motivated three largely separate lines of inquiry: (i)
positive-operator-valued-measure (POVM) formulations of time-of-arrival and
time-of-event
observables~\cite{Kijowski1974,Allcock1969,Grot1996,MugaLeavensPalao1998,EgusquizaMuga2000,Brunetti2002,Yearsley2010,Busch1994}
(Supplementary Materials, Section~\ref{sec:S3_POVM}); (ii) continuous-variable time--frequency modal
frameworks in chronocyclic phase
space~\cite{Brecht2013,Zhang2014,Fabre2020,Fabre2022,Descamps2023a,Descamps2023b,Descamps2024a,Descamps2024b,Descamps2026,Yoon2026a}
(Supplementary Materials, Section~\ref{sec:S4_TF_vs_relational}); and (iii) relational quantum clocks, in
which time emerges operationally by conditioning a target system on a
phase-stable reference
frame~\cite{Page1983,Moreva2014,Giovannetti2015,CastroRuizGiacominiBrukner2017,Giacomini2019,Favalli2020,Maccone2020,Giacomini2021_Quantum,Gambini2022,MeierSchwarzhansErkerHuber2023}
(Supplementary Materials, Sections~\ref{sec:S1_superselection} and \ref{sec:S2_Pauli}). In all three
frameworks, however, the \emph{direction} of time evolution is fixed externally
rather than treated as an active quantum degree of freedom.

A compelling theoretical paradigm that promotes temporal orientation itself to a
two-level quantum variable was recently introduced by Bloch, who formulated the
concept of the \emph{time qubit}~\cite{Bloch2025}. In Bloch's construction, a
composite system is defined on $\mathcal{H}=\mathcal{H}_T\otimes\mathcal{H}_S$,
where $\mathcal{H}_S$ is a target physical system and $\mathcal{H}_T$ is a
two-dimensional control Hilbert space spanned by $\{\ket{+}_T,\ket{-}_T\}$
(denoted simply as $\{\ket{+},\ket{-}\}$ hereafter). Physically, the orthonormal
logical states $\ket{+}$ and $\ket{-}$ represent definite, opposite temporal
orientations---corresponding to forward and backward dynamical evolution of the
target system---and define the north and south poles of a temporal Bloch sphere
as the $\pm 1$ eigenstates of the Pauli operator
$\tau_z=\ket{+}\!\bra{+}-\ket{-}\!\bra{-}$. 

In Bloch's single-particle Mach--Zehnder model~\cite{Bloch2025}, a pre-existing particle is first split by a passive input beam splitter into two spatial paths encoding $\ket{+}$ and $\ket{-}$. These paths are associated with a pair of branch Hamiltonians on $\mathcal{H}_S$ that differ by the sign of a time-direction-determining interaction $\hat{V}$~\cite{Bloch2025},
\begin{equation}
\hat{H}_{\pm}=\hat{H}_0\pm\hat{V},
\label{eq:H_pm}
\end{equation}
where $\hat{H}_0$ denotes the time-symmetric background dynamics. The composite
dynamics then take the controlled-Hamiltonian form~\cite{Bloch2025}
\begin{equation}
\hat{H}_{\rm tot}=\ket{+}\!\bra{+}\otimes\hat{H}_+ + \ket{-}\!\bra{-}\otimes\hat{H}_-
= I_T\otimes\hat{H}_0+\tau_z\otimes\hat{V},
\label{eq:H_tot}
\end{equation}
so that preparing the time qubit in $\ket{+}$ or $\ket{-}$ selectively evolves
$\mathcal{H}_S$ under $\hat{H}_+$ or $\hat{H}_-$. Conversely, when the time qubit is initialized in an equatorial superposition $(\ket{+}\pm\ket{-})/\sqrt{2}$ and the two branches are recombined at an output beam splitter, Bloch's general formulation evaluates the target evolution starting from an initial system state $\ket{\psi_0} \in \mathcal{H}_S$ (such as an arbitrary input state). This recombination projects the target evolution onto time-even and time-odd conditional Kraus operators~\cite{Bloch2025},
\begin{equation}
\hat{U}_{\mathrm{even}}(t)=\frac{1}{2}\!\left(e^{-\frac{i}{\hbar}\hat{H}_+ t}+e^{-\frac{i}{\hbar}\hat{H}_- t}\right), \qquad
\hat{U}_{\mathrm{odd}}(t)=\frac{1}{2}\!\left(e^{-\frac{i}{\hbar}\hat{H}_+ t}-e^{-\frac{i}{\hbar}\hat{H}_- t}\right),
\label{eq:Kraus_even_odd}
\end{equation}
which satisfy the trace-preserving completeness relation
$\hat{U}_{\mathrm{even}}^\dagger\hat{U}_{\mathrm{even}}+\hat{U}_{\mathrm{odd}}^\dagger\hat{U}_{\mathrm{odd}}=I_S$
(Supplementary Materials, Section~\ref{sec:S5_3_HL_and_Loschmidt}). Consequently, the resulting port-resolved detection probabilities
$P_{\mathrm{even/odd}}(t)=\bra{\psi_0}\hat{U}_{\mathrm{even/odd}}^\dagger(t)\hat{U}_{\mathrm{even/odd}}(t)\ket{\psi_0}$
manifest as \emph{time-parity fringes}~\cite{Bloch2025}.

While Bloch illustrated this framework through a conceptual spin-$1/2$ Larmor clock entering a conventional Mach--Zehnder interferometer with opposite magnetic fields $\pm\bm{B}$ (and its relativistic analog in the Dirac equation), Bloch's proposal has remained a theoretical gedankenexperiment restricted to passive pre-existing particles~\cite{Bloch2025}. To realize this framework dynamically, we replace Bloch's passive single-particle setup with an active quantum-optical architecture. Here, the particle input and signal beam splitter are replaced by two quantum-vacuum-seeded pair-creation sources embedded directly in the interferometer arms~\cite{Lee2018,Yoon2021}. This active setup raises three fundamental physical questions that cannot be addressed within passive spin models: 

First, can opposite Hamiltonian evolutions $\hat{H}_{\pm}=\hat{H}_0\pm\hat{V}$ and their coherent superposition be generated \emph{simultaneously with the creation of a single photon out of the quantum vacuum} via nonlinear phase matching, without any pre-existing input signal photon or signal input beam splitter? Second, what microscopic physical mechanism governs the complementarity of the arrow of time---that is, the \emph{parametric crossover between a definite temporal orientation and a pure coherent superposition} of opposite temporal orientations (equatorial $\tau_{x,y}$ coherence tuned by the vacuum-seed ratio)? Third, how does a discrete path-encoded time qubit structure an infinite-dimensional continuous-variable electromagnetic time--frequency wavepacket~\cite{Dayan2007,Walmsley2009,Eckstein2011,Brecht2013,Fabre2020,Fabre2022,Yoon2026a} alongside a Page--Wootters relational clock~\cite{Page1983,Maccone2020,Favalli2020}, and what \emph{direct first-order physical observables} does it measure beyond conventional Franson~\cite{Franson1989,Zhang2014} or time-bin interferometers?

\begin{figure}[tbh]
\centering
\includegraphics[width=\linewidth]{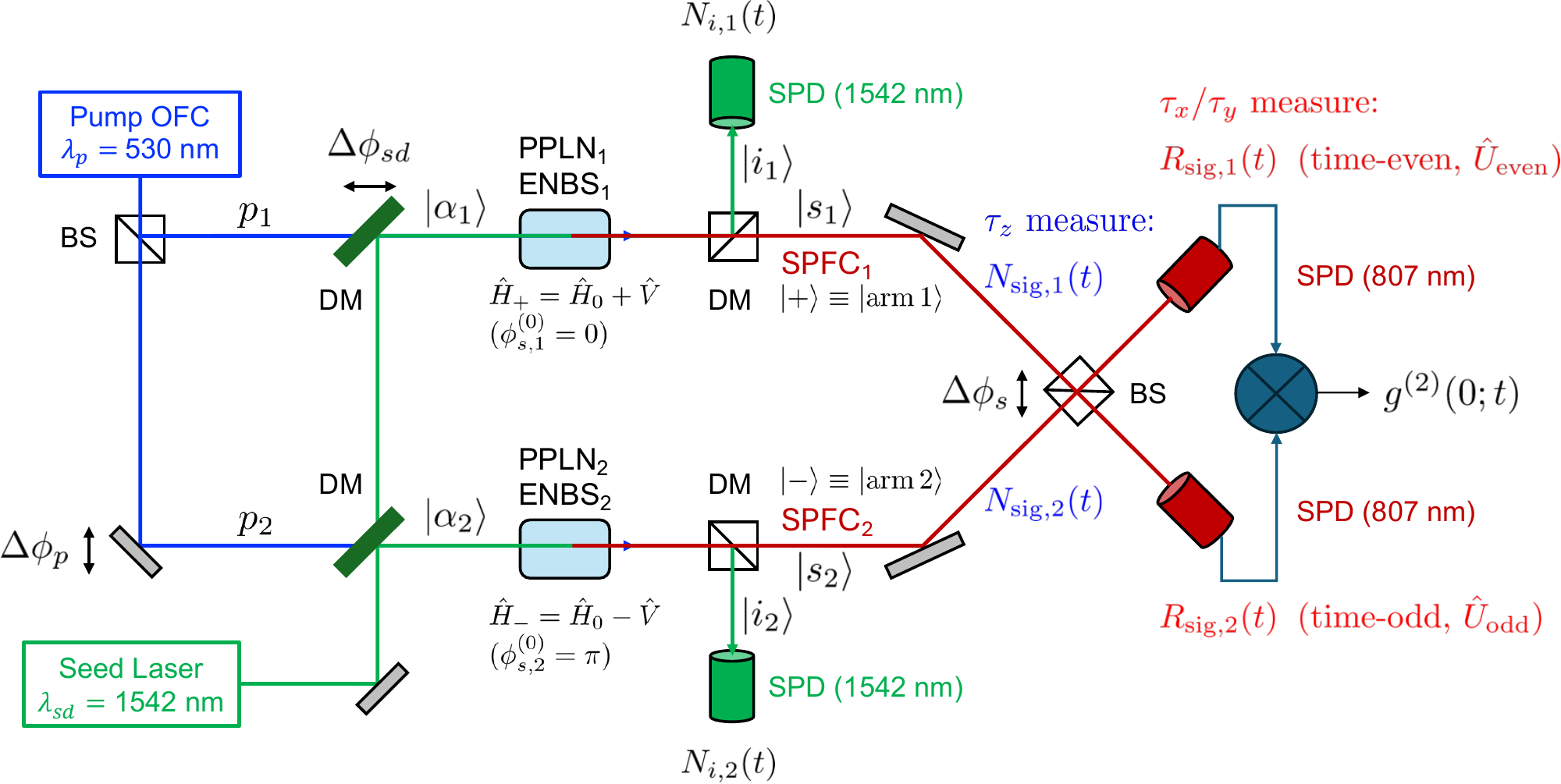}
\caption{
\textbf{Hybrid nonlinear--linear interferometer architecture of the single-photon time qubit.}
Two parallel, coherently seeded sources ($\mathrm{ENBS}_1$ and $\mathrm{ENBS}_2$ in $\mathrm{PPLN}_1$ and $\mathrm{PPLN}_2$ waveguides) replace the input beam splitter of a conventional Mach--Zehnder interferometer to define the logical time-qubit paths $\ket{+}\equiv\ket{\mathrm{arm\ 1}}$ and $\ket{-}\equiv\ket{\mathrm{arm\ 2}}$.
Driven by a common pump field ($\lambda_p=530$~nm; arms $p_1, p_2$ with relative phase $\Delta\phi_p$) and a shared monochromatic idler seed ($\lambda_{sd}=1542$~nm; coherent states $\ket{\alpha_1}, \ket{\alpha_2}$ with relative phase $\Delta\phi_{sd}$), setting the phase-matched signal creation phases to $\phi_{s,1}^{(0)}=0$ and $\phi_{s,2}^{(0)}=\pi$ realizes the opposite Hamiltonian branches $\hat{H}_{\pm}=\hat{H}_0\pm\hat{V}$ on the generated $807$~nm signal modes $\ket{s_1},\ket{s_2}$ ($\mathrm{SPFC}_1,\mathrm{SPFC}_2$), while the $1542$~nm idler modes $\ket{i_1},\ket{i_2}$ ($N_{i,1}(t), N_{i,2}(t)$) are separated without recombination.
Removing the output BS yields branch-resolved photon numbers $N_{\mathrm{sig},1}(t)$ and $N_{\mathrm{sig},2}(t)$ ($\tau_z$ readout), whereas balanced recombination with readout phase $\Delta\phi_s$ projects onto the time-even ($\hat{U}_{\mathrm{even}}$) and time-odd ($\hat{U}_{\mathrm{odd}}$) ports $R_{\mathrm{sig},1}(t)$ and $R_{\mathrm{sig},2}(t)$ ($\tau_x/\tau_y$ readout), with $g^{(2)}(0;t)$ certifying the single-photon regime.
OFC, optical frequency comb; BS, beam splitter; DM, dichroic mirror; PPLN, periodically poled lithium niobate; ENBS, entangled nonlinear biphoton source; SPFC, single-photon frequency comb; SPD, single-photon detector.}\label{fig:fig1}
\end{figure}

Here we answer these questions by formulating the quantum-optical theory of a
\emph{quantum-vacuum-seeded single-photon time qubit}. We experimentally probe
and validate its predictions using the hybrid nonlinear--linear
\emph{entangled nonlinear biphoton source} (ENBS) interferometry platform
introduced in Refs.~\cite{Lee2018,Yoon2021} and extended in
Refs.~\cite{Yoon2025,Yoon2026a,Yoon2026b}. Unlike a standard Mach--Zehnder
interferometer that splits an incident single photon at an input beam splitter,
our hybrid interferometer (Fig.~\ref{fig:fig1}) replaces the signal input beam
splitter with two parallel periodically poled lithium niobate (PPLN) waveguides.
These waveguides are driven by a common optical-frequency-comb (OFC) pump and a shared monochromatic
idler seed. They generate phase-coherent single-photon frequency-comb (SPFC)
supermodes directly from the quantum vacuum prior to their recombination at a
single output beam splitter. Specifically, we establish four advances that
connect quantum-vacuum control and complementarity of the arrow of time to
experiment.

\emph{First}, under three-wave phase matching in the two parallel
PPLN waveguides (Supplementary Materials, Sections~\ref{sec:S5_ENBS} and \ref{sec:S6_comparison}),
decomposing the coherently seeded $\chi^{(2)}$ interaction separates the
dynamics into a phase-insensitive spontaneous two-mode squeezing background $\hat{H}_0$
and a phase-sensitive stimulated displacement generator
$\hat{V}$~\cite{Yoon2026a,Yoon2026b}. Imprinting a relative $\pi$ pump--seed
creation phase between the two waveguides inverts $\hat{V}\to -\hat{V}$ while
leaving $\hat{H}_0$ invariant. This drives the collective SPFC time--frequency
supermode---which is defined in chronocyclic time--frequency $(t,\omega)$ phase
space rather than conventional mechanical position--momentum $(x,p)$ phase space
(Supplementary Materials, Section~\ref{sec:S4_TF_vs_relational})---with opposite displacement velocities
$\pm g_0$. Consequently, the stimulated evolution in the second arm is mathematically
identical to backward time evolution in the first arm,
$\hat{U}_{\mathrm{stim},-}(t)=\hat{U}_{\mathrm{stim},+}(-t)$
(Supplementary Materials, Section~\ref{sec:S5_ENBS}).

\emph{Second}, solving for the joint path--signal--idler state
(Supplementary Materials, Section~\ref{sec:S5_2_state_derivation}) reveals that spontaneous pair creation
excites orthogonal idler vacuum fluctuations that record complete which-branch
(temporal-direction) information. Conversely, stimulated emission into the shared
coherent seed $\ket{\alpha}$ prepares a coherent equatorial superposition of
$\ket{+}$ and $\ket{-}$ across the parallel arms, erasing which-branch
information without either a signal input beam splitter or the cascaded physical
idler alignment through both crystals required in unseeded induced-coherence
interferometers~\cite{Zou1991,Lemos2014}. This yields an exact wave--particle
complementarity law $|\bm{r}|=\mathcal{V}=|\alpha|^2/(1+|\alpha|^2)$ governing
the purity of the arrow of time~\cite{Lee2018,Yoon2021,Yoon2026b}.

\emph{Third}, building on our experimental phase-scan and visibility
data~\cite{Lee2018,Yoon2026b}, we demonstrate universal single-qubit
$\mathrm{SU}(2)$ state preparation and measurement---including non-Clifford
$T=R_Z(\pi/4)$ gates and postselection-free first-order tomography. We further prove
that the resulting time-parity fringes directly measure the complex
\emph{Loschmidt time-reversal echo amplitude}
$\mathcal{L}(t)=\bra{\psi_0}\hat{U}_-^\dagger(t)\hat{U}_+(t)\ket{\psi_0}$ without
coincidence postselection (Supplementary Materials, Section~\ref{sec:S5_3_HL_and_Loschmidt}).

\emph{Finally}, because the target system $\mathcal{H}_S$ is a $360$-mode SPFC
wavepacket anchored to a phase-locked optical frequency comb, our architecture
mathematically unifies Bloch's discrete time qubit ($\mathcal{H}_T$)~\cite{Bloch2025},
Page--Wootters relational clock conditioning ($\hat{\Pi}_C(t_c)$,
Supplementary Materials, Sections~\ref{sec:S1_superselection}--\ref{sec:S3_POVM})~\cite{Page1983,Maccone2020},
and chronocyclic Wigner time--frequency phase space ($W(t,\omega)$,
Supplementary Materials, Section~\ref{sec:S4_TF_vs_relational})~\cite{Brecht2013,Fabre2020,Fabre2022,Yoon2026a}
within a single photonic platform.

The remainder of this paper is structured as follows. In the following sections,
we detail the hybrid interferometric platform and its rigorous operator mapping
onto Bloch's time-qubit Hamiltonian. We then demonstrate phase-stable
superpositions of opposite temporal orientations, establish universal
single-qubit $\mathrm{SU}(2)$ control and postselection-free first-order
tomography, and discuss broader applications for multi-qubit temporal photonics
and operational tests of quantum time.

\section*{Results}

\subsection*{Phase-matched operator mapping and quantum-vacuum complementarity of the time qubit}

Figure~\ref{fig:fig1} illustrates the architecture of the single-photon time qubit. Two identical PPLN-waveguide ENBS units ($\mathrm{PPLN}_1$ and $\mathrm{PPLN}_2$) are placed in arms $j\in\{1,2\}$ of a Mach--Zehnder interferometer, defining the logical basis $\ket{+}\equiv\ket{\mathrm{arm\ 1}}$ and $\ket{-}\equiv\ket{\mathrm{arm\ 2}}$ of $\mathcal{H}_T$. Both waveguides are pumped by a common pulsed OFC field ($\lambda_p=530$~nm, repetition rate $f_{\mathrm{rep}}=250$~MHz, pump phases $\phi_{p,j}$) and seeded collinearly by a shared continuous-wave monochromatic idler laser ($\lambda_{sd}=1542$~nm, coherent amplitudes $\alpha_j=|\alpha_j|e^{i\phi_{\mathrm{sd},j}}$), generating broadband SPFC signal fields centered at $\lambda_s=807$~nm.

To establish the microscopic correspondence with Eqs.~\eqref{eq:H_pm} and \eqref{eq:H_tot} and clarify the exact phase-matching relation in Fig.~\ref{fig:fig1} (derived in Supplementary Materials, Section~\ref{sec:S5_1_Hamiltonian}), decomposing the coherently seeded $\chi^{(2)}$ interaction separates the interaction-picture Hamiltonian in $\mathrm{PPLN}_j$ into a phase-insensitive spontaneous two-mode squeezing background $\hat{H}_0$ and a phase-sensitive stimulated displacement generator $\hat{V}$ [detailed in Supplementary Materials, Section~\ref{sec:S5_1_Hamiltonian}, Supplementary Eqs.~(S18)--(S20)]~\cite{Yoon2026a,Yoon2026b}. Specifically, under symmetric pump splitting ($\Delta\phi_p=0$) and a controlled $\pi$ phase inversion ($\Delta\phi_{sd}=\pi$) yielding $g_1 = +g_0$ and $g_2 = -g_0$, mapping the local supermodes onto the isomorphic target modes yields Bloch's controlled Hamiltonian Eqs.~\eqref{eq:H_pm} and \eqref{eq:H_tot} with $\hat{V} = i\hbar(g_0 \hat{A}_{\mathrm{SM}}^\dagger - g_0^* \hat{A}_{\mathrm{SM}})$ [Supplementary Eq.~(S20)].

In our hybrid ENBS interferometer, the optical analog of Larmor precession is the coherent phase-space displacement of the SPFC signal supermode $\hat{A}_{\mathrm{SM}}$ in the interaction picture of $\hat{H}_0$, which unfolds natively in chronocyclic time--frequency phase space rather than mechanical position--momentum phase space (as detailed in Supplementary Materials, Section~\ref{sec:S4_TF_vs_relational}). Under the stimulated generator $\pm\hat{V}$, the Heisenberg equation of motion for the signal supermode amplitude gives displacement velocities $\pm g_0$ [Supplementary Eq.~(S43)]. Because $\hat{V}$ and time $t$ enter the unitary operator $\hat{U}_{\mathrm{stim},\pm}(t) = \exp(\mp i\hat{V}t/\hbar)$ through their product, inverting the creation phase by $\pi$ ($\hat{V}\to -\hat{V}$) is mathematically identical to time-reversal $\hat{U}_{\mathrm{stim},-}(t) = \hat{U}_{\mathrm{stim},+}(-t)$ [Supplementary Eq.~(S21)].

Crucially, unlike Bloch's spin interferometer~\cite{Bloch2025}, the coexistence of $\hat{H}_0$ (spontaneous two-mode squeezing background) and $\hat{V}$ (stimulated displacement) introduces a tunable quantum-vacuum mechanism that controls the purity of the temporal orientation (Supplementary Materials, Section~\ref{sec:S5_2_state_derivation}). Evolving the initial vacuum state produces the joint path--signal--idler state [Supplementary Eq.~(S22)], revealing an exact wave--particle complementarity governing the arrow of time: spontaneous idler vacuum fluctuations record complete which-branch ($\tau_z$) information, whereas stimulated emission into the shared coherent seed $\ket{\alpha}$ erases it without requiring idler alignment or postselection (Supplementary Materials, Sections~\ref{sec:S1_superselection} and \ref{sec:S5_2_state_derivation}). This yields the exact complementarity law for the temporal Bloch vector length and fringe visibility:
\begin{equation}
|\bm{r}| = \mathcal{V} = \frac{|\alpha|^2}{1+|\alpha|^2},
\label{eq:V_scaling_main}
\end{equation}
where the denominator $1+|\alpha|^2$ explicitly reflects the spontaneous background ($1$) and the stimulated contribution ($|\alpha|^2$).

\subsection*{Operational distinction from Franson, time-bin, and passive spin interferometry}

To situate our platform within the broader landscape of quantum interferometry, Table~\ref{tab:comparison} compares the ENBS time qubit against Franson time--energy interferometry~\cite{Franson1989,Zhang2014}, conventional time-bin qubits~\cite{Brendel1999}, unseeded Zou--Wang--Mandel (ZWM) induced coherence~\cite{Zou1991,Lemos2014}, and Bloch's passive spin-clock proposal~\cite{Bloch2025}. Analyzing these protocols side-by-side highlights the distinct operational advantages of our ENBS architecture. 

Unlike \textit{Franson interferometry}, which generates nonlocal correlations using unbalanced paths but relies strictly on fourth-order coincidence detection behind remote interferometers (exhibiting zero first-order visibility $\mathcal{V}^{(1)}=0$ and requiring severe coincidence postselection)~\cite{Franson1989,Zhang2014}, our ENBS time qubit operates via first-order temporal coherence, yielding high-visibility interference fringes directly in port-resolved single-photon counting rates without any postselection. Similarly, compared to \textit{conventional time-bin qubits} that encode information in discrete arrival slots ($\alpha\ket{t_{\mathrm{early}}} + \beta\ket{t_{\mathrm{late}}}$) using passive optical delay lines and require time-resolved single-photon gating (incurring a mandatory $50\%$ postselection loss)~\cite{Brendel1999}, our architecture encodes temporal orientation in active Hamiltonian branches $\hat{H}_{\pm}$ during vacuum pair creation rather than passive delay slots.

While \textit{unseeded Zou--Wang--Mandel (ZWM) induced coherence} demonstrates first-order signal interference from a source-crystal path superposition $(\ket{\mathrm{arm\ 1}}_s + e^{i\phi_i}\ket{\mathrm{arm\ 2}}_s)\ket{i}$~\cite{Zou1991,Lemos2014}, it requires physical cascaded alignment and spatial overlap of the idler modes ($\{i_1, i_2\} \equiv i$) through the second crystal, restricting visibility by idler transmission losses ($\mathcal{V} \le |t_i|$). By contrast, our ENBS platform achieves postselection-free first-order interference using a shared monochromatic coherent seed $\ket{\alpha}$ that drives stimulated emission without ever physically overlapping or detecting the idler beams. Furthermore, in contrast to \textit{Bloch's passive spin-clock model}~\cite{Bloch2025}, which treats the time qubit as a pre-existing 2D spin-$1/2$ particle precessing passively under external magnetic fields ($\hat{H}_{\pm} = \frac{\hat{p}^2}{2m} \mp \frac{g\mu_B}{2}\bm{\sigma}\cdot\bm{B}$), our optical implementation generates the temporal orientation dynamically from the quantum vacuum via phase-matched $\chi^{(2)}$ parametric down-conversion ($\phi_{s,j}^{(0)} = \phi_{p,j} - \phi_{\mathrm{sd},j}$), targets an infinite-dimensional multi-mode SPFC bosonic field, and analytically governs temporal coherence through quantum-vacuum wave--particle complementarity shown in Eq.~\eqref{eq:V_scaling_main}.

\begin{table}[tbh]
\centering
\caption{\textbf{Operational comparison of temporal interferometry protocols, Bloch's spin-clock proposal, and the ENBS single-photon time qubit.}}
\label{tab:comparison}
\footnotesize
\renewcommand{\arraystretch}{1.3}
\begin{tabular}{>{\raggedright\arraybackslash}p{2.55cm} >{\raggedright\arraybackslash}p{3.0cm} >{\raggedright\arraybackslash}p{3.05cm} >{\raggedright\arraybackslash}p{3.0cm} >{\raggedright\arraybackslash}p{3.1cm}}
\toprule
\textbf{Protocol / Framework} & \textbf{Target Space $\mathcal{H}_S$ \& Encoding Basis} & \textbf{Physical Mechanism of Superposition} & \textbf{Measured Observable \& Postselection} & \textbf{Distinctive Physical Signature} \\
\midrule
\textbf{Franson Interferometry}~\cite{Franson1989,Zhang2014} & Two-photon emission time: $\ket{t_{\mathrm{early}},t_{\mathrm{early}}} + e^{i(\phi_s+\phi_i)}\ket{t_{\mathrm{late}},t_{\mathrm{late}}}$ & Passive unbalanced MZIs ($\Delta T \gg \tau_{\mathrm{coh}}$) applied \emph{after} pair creation & Two-photon coincidence rate $R_{si}^{(2)}$; requires temporal postselection of zero-delay peak & Nonlocal fourth-order fringes ($\mathcal{V}^{(2)}>0$); strictly zero first-order fringes ($\mathcal{V}^{(1)}=0$) \\
\addlinespace[3pt]
\textbf{Conventional Time-Bin Qubit}~\cite{Brendel1999} & Discrete arrival slots of single photon: $\alpha\ket{t_{\mathrm{early}}} + \beta e^{i\phi}\ket{t_{\mathrm{late}}}$ & Passive optical delay lines before transmission and at analyzer & Time-resolved gating of central arrival slot ($50\%$ postselection loss) & Three-peak arrival histogram; interference confined to central time slot \\
\addlinespace[3pt]
\textbf{Unseeded Induced Coherence}~\cite{Zou1991,Lemos2014} & Source-crystal path of SPDC pair: $(\ket{\mathrm{arm\ 1}}_s + e^{i\phi_i}\ket{\mathrm{arm\ 2}}_s)\ket{i}$ & Cascaded physical alignment of idler modes $\{i_1, i_2\} \equiv i$ through second crystal & First-order signal counts $R_{\mathrm{sig}}$; requires physical idler overlap & First-order visibility bounded by idler transmission loss ($\mathcal{V}\le |t_i|$) \\
\addlinespace[3pt]
\textbf{Bloch Spin-Clock Time Qubit}~\cite{Bloch2025} & 2D spin-$1/2$ clock precessing under $\hat{H}_{\pm}=\frac{\hat{p}^2}{2m}\mp\frac{g\mu_B}{2}\bm{\sigma}\cdot\bm{B}$ & Passive Zeeman precession of a pre-existing particle in opposite fields $\pm\bm{B}$ & State-selective spin-port detection $P_{D_{1,2}}$; no vacuum generation & Spin Larmor time-parity fringes $\cos^2(\Omega T/2)$; qualitative decoherence \\
\addlinespace[3pt]
\textbf{ENBS Photonic Time Qubit (This Work)} & $N$-mode SPFC bosonic supermode under $\hat{H}_{\pm}=\hat{H}_0\pm\hat{V}$ [Supplementary Materials, Section~\ref{sec:S5_1_Hamiltonian}, Supplementary Eqs.~(S18)--(S20)] & Phase-matched creation phase $\phi_{s,j}^{(0)}=\phi_{p,j}-\phi_{\mathrm{sd},j}$ with $\Delta\phi_s^{(0)}=\phi_{s,1}^{(0)}-\phi_{s,2}^{(0)}=\pm\pi$ & Port-resolved \textbf{first-order} counts $R_{\mathrm{sig},1/2}$; \textbf{no postselection} or idler overlap & Exact vacuum complementarity $|\bm{r}|=\frac{|\alpha|^2}{1+|\alpha|^2}$; direct Loschmidt echo \& PW clock conditioning \\
\bottomrule
\end{tabular}
\end{table}

As summarized in Table~\ref{tab:comparison}, our architecture unites the conceptual framework of time qubits with a scalable, postselection-free photonic platform that requires neither coincidence counting, bin-gating, nor idler-mode overlap.

\begin{figure*}[tbh]
\centering
\includegraphics[width=0.98\textwidth]{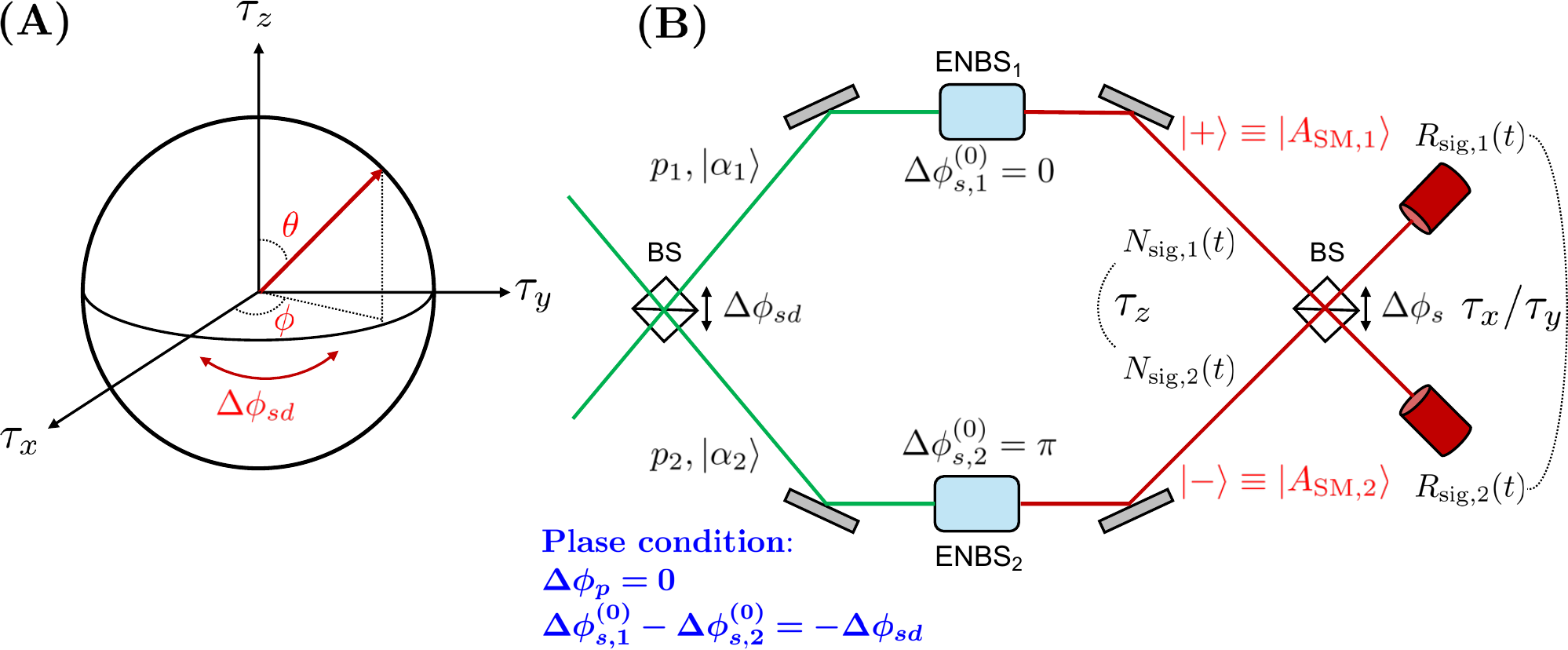}
\caption{\textbf{Phase-stable interference and temporal Bloch-sphere representation of the single-photon time qubit.} 
(\textbf{A}) Temporal Bloch-sphere representation of the time qubit $\mathcal{H}_T$. The logical states $\ket{+}$ and $\ket{-}$ (poles along the $\tau_z$ axis) correspond to opposite temporal orientations governed by the creation-stage Hamiltonian branches $\hat{H}_{\pm} = \hat{H}_0 \pm \hat{V}$ [Supplementary Materials, Section~\ref{sec:S5_1_Hamiltonian}, Supplementary Eqs.~(S18)--(S20)]. Varying the creation-stage phase difference $\Delta\phi_{sd}$ (under the phase condition $\Delta\phi_p=0$ and $\Delta\phi_{s,1}^{(0)} - \Delta\phi_{s,2}^{(0)} = -\Delta\phi_{sd}$) prepares coherent superposition states $(\ket{+} + e^{i\Delta\phi_{sd}}\ket{-})/\sqrt{2}$, implementing $R_Z$ rotations about the $\tau_z$ axis with azimuthal angle $\phi$. 
(\textbf{B}) Interferometric ENBS readout architecture. Two parallel periodically poled lithium niobate ($\mathrm{PPLN}_1$ and $\mathrm{PPLN}_2$) waveguides embedded in a Mach--Zehnder interferometer generate SPFC supermodes $\hat{A}_{\mathrm{SM},1}$ and $\hat{A}_{\mathrm{SM},2}$ seeded by $\ket{\alpha_1}$ and $\ket{\alpha_2}$. Removing the output beam splitter yields branch-resolved detection of the signal photon numbers $N_{\mathrm{sig},j}(t)$, corresponding to a projective measurement of $\tau_z$, whereas balanced recombination with a tunable post-creation signal phase $\Delta\phi_s$ rotates the detector eigenmodes on the equator, enabling access to the $\tau_x$ and $\tau_y$ time-parity observables and direct first-order measurement of the Loschmidt echo amplitude via port-resolved counting rates $R_{\mathrm{sig},1/2}(t)$.}
\label{fig:fig2}
\end{figure*}
\FloatBarrier

\subsection*{Time-parity fringes and first-order Loschmidt-echo quantum simulation}
\label{sec:time_parity_fringes}

We next demonstrate how coherent superpositions of opposite temporal orientations are prepared during nonlinear pair generation and used to measure the Loschmidt time-reversal echo amplitude through first-order temporal interference (detailed in Supplementary Materials, Section~\ref{sec:S5_3_HL_and_Loschmidt}). Figure~\ref{fig:fig2} illustrates how the creation-stage pump--seed phase difference ($\Delta\phi_{sd}$ under the phase condition $\Delta\phi_p=0$ and $\phi_{s,1}^{(0)}-\phi_{s,2}^{(0)} = -\Delta\phi_{sd}$, referenced in Supplementary Materials, Section~\ref{sec:S6_1_phase_variables}) and the post-creation interferometric signal-phase shift $\Delta\phi_s$ provide independent control over state preparation and measurement.

Varying the relative seed phase $\Delta\phi_{sd}$ (with $\Delta\phi_p=0$) modulates the creation-stage phase difference during nonlinear pair generation, preparing the stimulated time-qubit state
\begin{equation}
\ket{\psi_T} = \frac{1}{\sqrt{2}}\!\left(\ket{+} + e^{i\Delta\phi_{sd}}\ket{-}\right),
\label{eq:psi_T}
\end{equation}
which rotates the temporal Bloch vector azimuthally around the $\tau_z$ axis [Fig.~\ref{fig:fig2}(A)]. Recombining the SPFC signal fields at the output beam splitter with a tunable post-creation signal phase $\Delta\phi_s$ projects onto the generalized time-parity Kraus operators $\hat{U}_{\mathrm{even/odd}}(t;\Delta\phi_s)$ [Supplementary Materials, Section~\ref{sec:S5_3_HL_and_Loschmidt}, Supplementary Eq.~(S40)]. 

Crucially, as derived explicitly in Supplementary Materials, Section~\ref{sec:S5_3_HL_and_Loschmidt} [Supplementary Eqs.~(S40)--(S42)], we uncover a direct operational link between these optical time-parity fringes and the \textit{Loschmidt time-reversal echo} $\mathcal{L}(t) = \bra{\psi_0}\hat{U}_-^\dagger(t)\hat{U}_+(t)\ket{\psi_0}$. Because the interferometer path $\mathcal{H}_T$ acts natively as the control ancilla during photon creation, the normalized first-order counting-rate contrast between the two output ports directly measures the real and imaginary quadratures of the conditional Loschmidt overlap [Eq.~\eqref{eq:Loschmidt_main}]:
\begin{equation}
\frac{R_{\mathrm{sig,1}}(t)-R_{\mathrm{sig,2}}(t)}{R_{\mathrm{sig,1}}(t)+R_{\mathrm{sig,2}}(t)}
= \mathcal{V}(|\alpha|)\,\cos(\Delta\phi_{sd}+\Delta\phi_s)
= \mathcal{V}(|\alpha|)\,\Re\!\left[e^{i\Delta\phi_s}\frac{\bra{\psi_+(t)}\ket{\psi_-(t)}}{\|\ket{\psi_+(t)}\|\,\|\ket{\psi_-(t)}\|}\right],
\label{eq:Loschmidt_main}
\end{equation}
providing direct, postselection-free access to the real and imaginary parts of the time-reversal echo amplitude (including simulating time-reversal asymmetries in Unruh--DeWitt detector models~\cite{Yoon2025}).

\begin{figure}[t]
\centering
\includegraphics[width=0.98\linewidth]{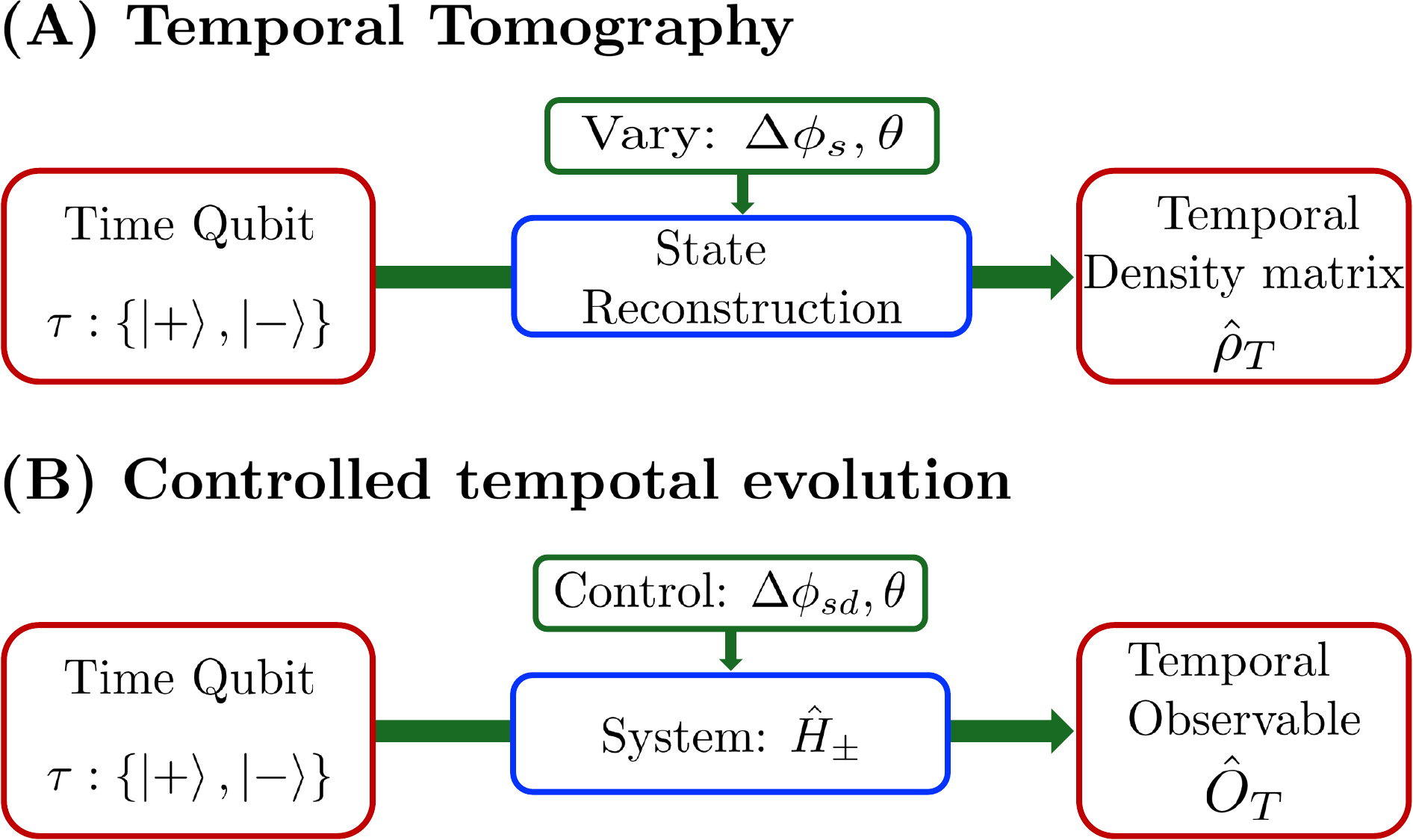}
\caption{
\textbf{Operational architecture for temporal state tomography and controlled evolution.}
(\textbf{A}) Temporal Tomography workflow: starting from the time qubit $\mathcal{H}_T = \mathrm{span}\{\ket{+},\ket{-}\}$, varying the interferometric readout phase $\Delta\phi_s$ and seed-amplitude ratio $\theta$ reconstructs the full temporal density matrix $\hat{\rho}_T$ from port-resolved single-photon counting rates via state reconstruction (Supplementary Materials, Section~\ref{sec:S6_2_measurement_basis}).
(\textbf{B}) Controlled Temporal Evolution workflow: actively tuning the creation-stage pump--seed phase difference $\Delta\phi_{sd}$ and seed-amplitude ratio $\theta$ programs the Hamiltonian branches $\hat{H}_{\pm} = \hat{H}_0 \pm \hat{V}$ [Supplementary Materials, Section~\ref{sec:S5_1_Hamiltonian}, Supplementary Eqs.~(S18)--(S20)], enabling direct readout of generalized temporal observables $\hat{O}_T$ including Loschmidt-echo amplitude quadratures [Eq.~\eqref{eq:Loschmidt_main}].}
\label{fig:fig3}
\end{figure}

\subsection*{Universal single-qubit $\mathrm{SU}(2)$ control on the temporal Bloch sphere}
\label{sec:SU2_control}

\begin{figure*}[tbh]
\centering
\includegraphics[width=0.98\textwidth]{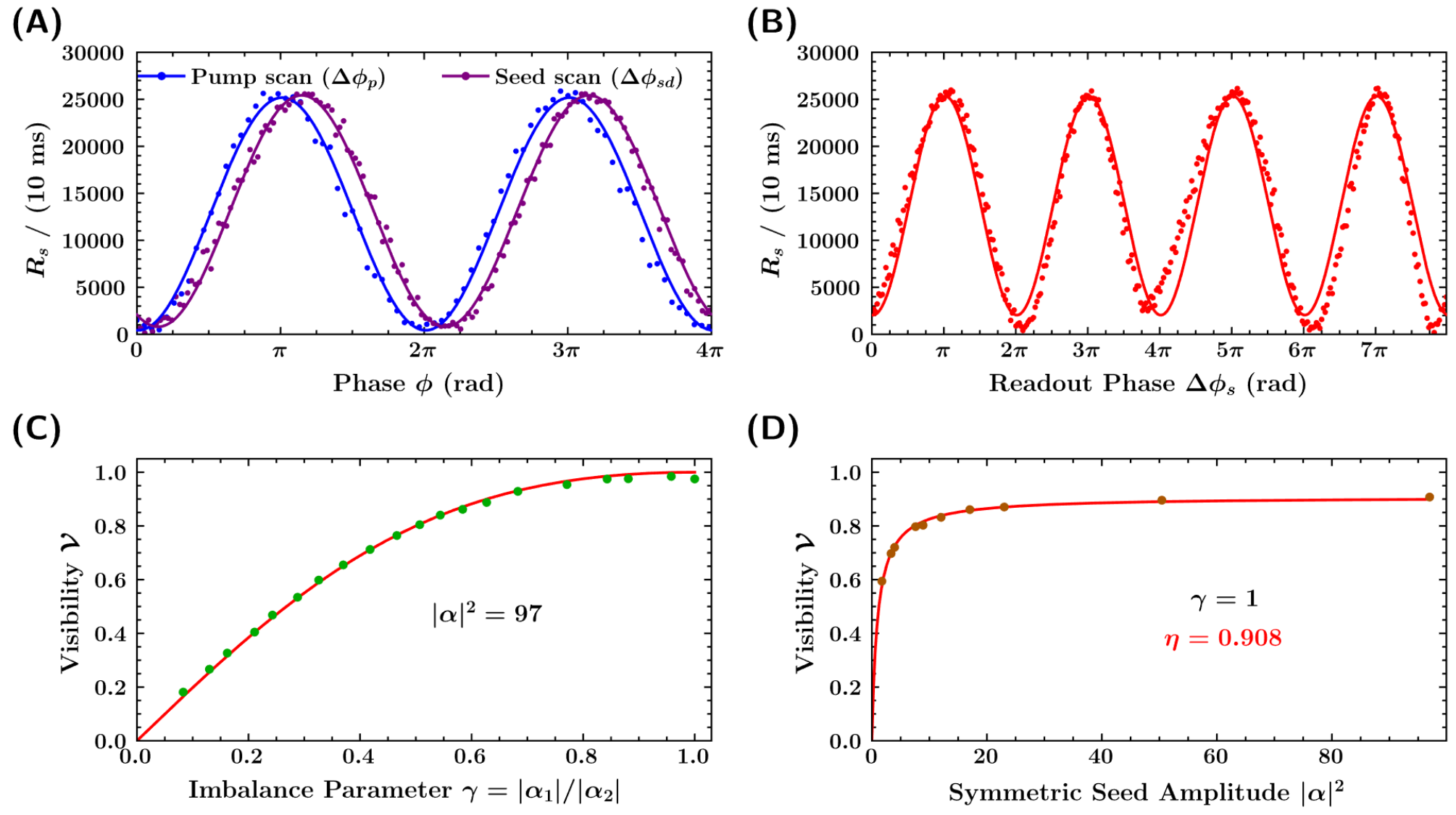}
\caption{\textbf{Experimental phase calibration, measurement-basis control, and visibility scaling of the single-photon time qubit.} 
(\textbf{A}) Single-photon signal counting rates $R_s$ (counts per 10\,ms) versus phase $\phi$ for open-loop pump-path ($\Delta\phi_p$, blue) and seed-path ($\Delta\phi_{sd}$, purple) scans, implementing azimuthal $R_Z(\phi)$ rotations on the temporal Bloch sphere. Solid curves represent sinusoidal fits.
(\textbf{B}) Readout-basis control via post-creation signal phase $\Delta\phi_s$ (red), scanning the detector eigenmodes on the equator between the $\tau_x$ and $\tau_y$ time-parity bases via even and odd output port supermode projections.
(\textbf{C}) Interference visibility $\mathcal{V}$ versus the coherent seed imbalance parameter $\gamma = |\alpha_1|/|\alpha_2|$ at fixed $|\alpha|^2 = 97$ (green circles), fitted by the theoretical polar-control relation $\mathcal{V} = 2\gamma/(1+\gamma^2)$ (red solid curve).
(\textbf{D}) Interference visibility $\mathcal{V}$ versus symmetric coherent seed amplitude $|\alpha|^2$ at $\gamma = 1$ (brown circles), fitted by the vacuum-complementarity scaling $\mathcal{V} = |\alpha|^2/(1+|\alpha|^2)$ with extracted mode overlap $\eta = 0.908$ (red solid curve). Panels (C) and (D) adapt data from Refs.~\cite{Lee2018,Yoon2026b}.}
\label{fig:fig4}
\end{figure*}

Going beyond Bloch's conceptual interferometer~\cite{Bloch2025}, we next formulate a complete set of universal single-qubit $\mathrm{SU}(2)$ state preparation and measurement (SPAM) operations on the temporal Bloch sphere and show that their core physical mechanisms are experimentally realized by our phase-resolved ENBS data~\cite{Lee2018,Yoon2021} (detailed in Supplementary Materials, Section~\ref{sec:S6_comparison}). Figure~\ref{fig:fig3} summarizes the operational workflows for temporal tomography and controlled Hamiltonian evolution, while Fig.~\ref{fig:fig4} presents the expanded four-panel experimental dataset---separating creation-stage azimuthal phase extractions [Fig.~\ref{fig:fig4}(A)], post-creation readout basis rotations [Fig.~\ref{fig:fig4}(B)], polar-angle visibility scaling versus imbalance parameter $\gamma$ [Fig.~\ref{fig:fig4}(C)], and vacuum-complementarity scaling versus symmetric seed amplitude $|\alpha|^2$ [Fig.~\ref{fig:fig4}(D)]---adapted from Refs.~\cite{Lee2018,Yoon2026b}.

\begin{figure*}[tbh]
\centering
\includegraphics[width=0.98\textwidth]{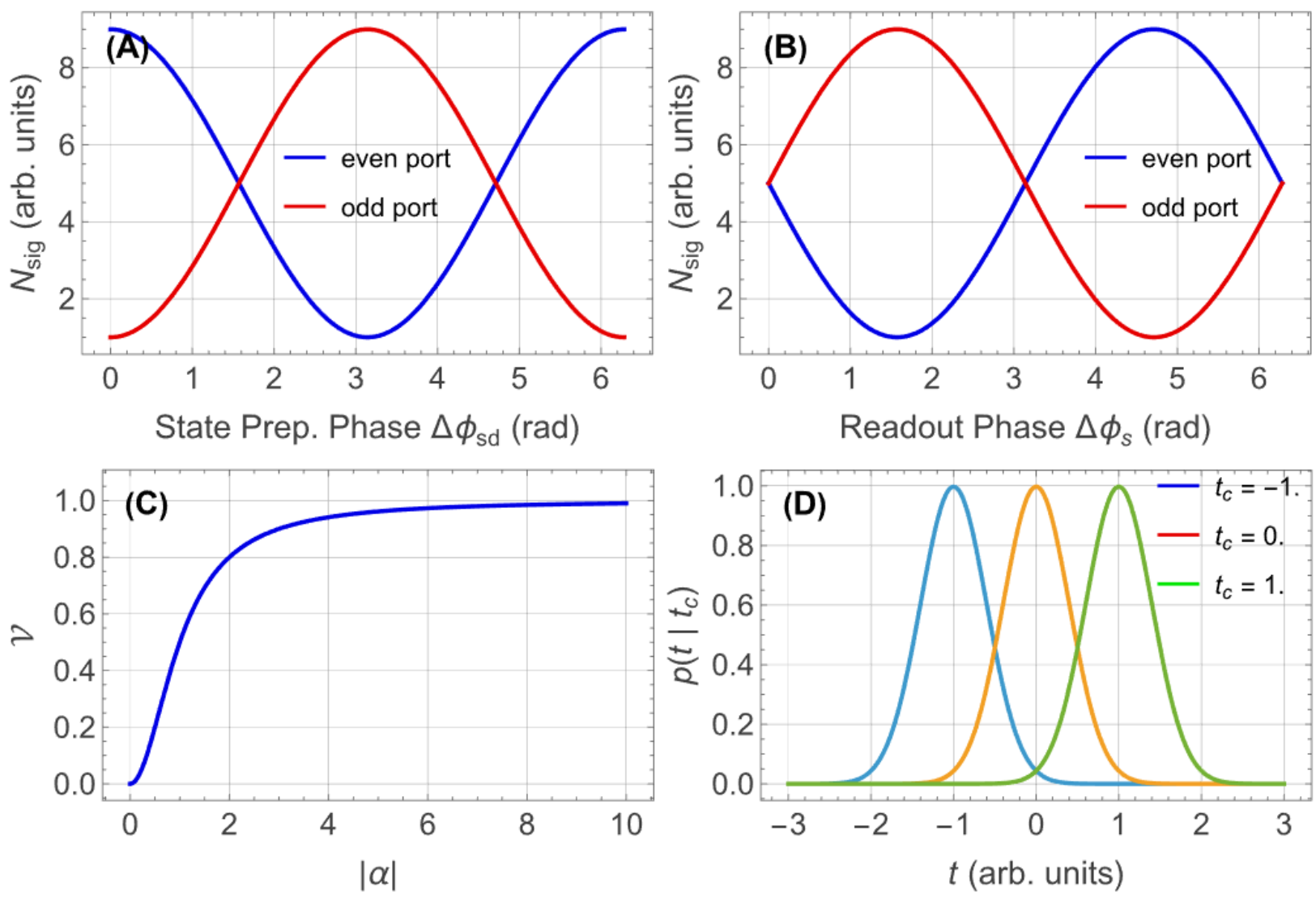}
\caption{\textbf{Theoretical time-qubit control, visibility, and relational clock conditioning under stabilized operation.} 
(\textbf{A}) Calculated port-resolved single-photon signal numbers $N_{\mathrm{sig}}$ at the even (blue) and odd (red) output ports as a function of the creation-stage seed phase $\Delta\phi_{sd}$, illustrating deterministic $R_Z(\Delta\phi_{sd})$ azimuthal rotations [Eq.~\eqref{eq:RZ}].
(\textbf{B}) Dependence of port-resolved signal outputs on the post-creation readout phase $\Delta\phi_s$, rotating the detector eigenmodes on the equator between the $\tau_x$ and $\tau_y$ time-parity bases via $R_Y(\Delta\phi_s)$ [Eq.~\eqref{eq:RY}].
(\textbf{C}) Analytical interference visibility $\mathcal{V} = |\alpha|^2/(1+|\alpha|^2)$ versus coherent seed amplitude $|\alpha|$ [Eq.~\eqref{eq:V_scaling_main}].
(\textbf{D}) Conditional temporal arrival-time probability distributions $p(t|t_c)$ of the SPFC signal wavepacket for three Page--Wootters relational clock settings $t_c = -1, 0, +1$ (in units of the transform-limited envelope width $\sigma_t$), derived from the covariant clock POVM and comb-conditioning map (Supplementary Materials, Sections~\ref{sec:S2_Pauli}--\ref{sec:S3_POVM}).}
\label{fig:fig5}
\end{figure*}

Universal single-qubit $\mathrm{SU}(2)$ control is achieved via three independent optical control knobs (Supplementary Materials, Section~\ref{sec:S6_comparison}):
\begin{enumerate}
\item \textbf{Azimuthal $R_Z(\phi)$ rotations (including non-Clifford $T$ gates):} Varying the creation-stage pump--seed phase difference $\phi = \Delta\phi_{sd} - \Delta\phi_p = \phi_{s,2}^{(0)} - \phi_{s,1}^{(0)} = -\Delta\phi_s^{(0)}$ implements continuous rotations about the $\tau_z$ axis [Fig.~\ref{fig:fig3}(B) and Supplementary Materials, Section~\ref{sec:S6_1_phase_variables}],
\begin{equation}
R_Z(\phi)=\exp\!\left(-i\frac{\phi}{2}\tau_z\right), \qquad \phi=\Delta\phi_{sd}\quad (\text{for }\Delta\phi_p=0).
\label{eq:RZ}
\end{equation}
As demonstrated experimentally by the open-loop phase-scan data in Fig.~\ref{fig:fig4}(A)---where independent scans of the pump path ($\Delta\phi_p$) and seed path ($\Delta\phi_{sd}$) produce $2\pi$-periodic sinusoidal fringes confirming $\phi_{s,j}^{(0)}=\phi_{p,j}-\phi_{\mathrm{sd},j}$ (Supplementary Materials, Section~\ref{sec:S6_1_phase_variables})---and modeled under closed-loop stabilization in Fig.~\ref{fig:fig5}(A), $\Delta\phi_{sd}$ is continuously tunable over multiple $2\pi$ cycles. Setting $\Delta\phi_{sd}=\pi$ implements the Pauli-$Z$ gate (inverting $\hat{V}\to -\hat{V}$), $\Delta\phi_{sd}=\pi/2$ implements the Clifford phase gate $S$, and $\Delta\phi_{sd}=\pi/4$ implements the \textbf{non-Clifford $\pi/8$ ($T$) gate} $R_Z(\pi/4)=\exp(-i\frac{\pi}{8}\tau_z)$.
\item \textbf{Polar $R_X(\theta)$ control via seed-amplitude imbalance:} Controlling the ratio of the coherent idler seed amplitudes $|\alpha_2|/|\alpha_1|$ between $\mathrm{PPLN}_1$ and $\mathrm{PPLN}_2$ tunes the relative stimulated pair-creation weights, setting the polar angle $\theta$ on the temporal Bloch sphere [Fig.~\ref{fig:fig3}(A,B), Fig.~\ref{fig:fig4}(C), and Supplementary Materials, Section~\ref{sec:S6_2_measurement_basis}] according to~\cite{Yoon2026b}
\begin{equation}
\tan\!\left(\frac{\theta}{2}\right)=\frac{|\alpha_2|}{|\alpha_1|}=\gamma^{-1}, \qquad R_X(\theta)=\exp\!\left(-i\frac{\theta}{2}\tau_x\right).
\label{eq:RX}
\end{equation}
Because $R_X(\theta)$ and $R_Z(\phi)$ do not commute ($[\tau_x,\tau_z]=-2i\tau_y\neq 0$), combining arbitrary $R_X(\theta)$ and $R_Z(\phi)$ (including the non-Clifford angle $\phi=\pi/4$) provides universal single-qubit $\mathrm{SU}(2)$ state preparation across the entire temporal Bloch sphere (with multi-qubit extensions discussed in Supplementary Materials, Section~\ref{sec:S7_extensions}).
\item \textbf{Readout basis rotation $R_Y(\beta)$:} At the output beam splitter, the post-creation signal-phase shift $\Delta\phi_s$ rotates the projective measurement axis [Fig.~\ref{fig:fig3}(A) and Supplementary Materials, Section~\ref{sec:S6_2_measurement_basis}] via
\begin{equation}
R_Y(\beta)=\exp\!\left(-i\frac{\beta}{2}\tau_y\right), \qquad \beta=\Delta\phi_s,
\label{eq:RY}
\end{equation}
continuously rotating the detector eigenmodes on the equator to project onto the even and odd output port supermodes $\frac{1}{\sqrt{2}}(\ket{+} \pm e^{i\Delta\phi_s}\ket{-})$ [Fig.~\ref{fig:fig4}(B)], which yields the $\tau_x$ quadrature at $\Delta\phi_s=0$ and the $\tau_y$ quadrature at $\Delta\phi_s=\pi/2$, as shown in Fig.~\ref{fig:fig5}(B).
\end{enumerate}
We distinguish here between \emph{universal single-qubit $\mathrm{SU}(2)$ SPAM control}---which is completely realized by the noncommuting continuous operations $\{R_Z(\Delta\phi_{sd}), R_X(\theta), R_Y(\Delta\phi_s)\}$---and multi-qubit circuit universality, which additionally requires two-qubit entangling gates between distinct time qubits (formulated in Supplementary Materials, Section~\ref{sec:S7_extensions})~\cite{Yoon2026a}.

Figure~\ref{fig:fig4}(D) shows the experimentally measured visibility from Ref.~\cite{Lee2018} as a function of the symmetric seed amplitude $|\alpha|^2$, while Fig.~\ref{fig:fig5}(C) displays the analytical prediction Eq.~\eqref{eq:V_scaling_main} (see Supplementary Materials, Section~\ref{sec:S6_3_visibility_and_budget} and Supplementary Eq.~(S49) for the quantitative extraction of $\eta_{\mathrm{mode}}$~\cite{Yoon2026b}), confirming the crossover from a maximally mixed temporal state in the spontaneous limit ($|\alpha|\ll 1$) to a pure temporal superposition in the stimulated regime ($|\alpha|\gg 1$).

\FloatBarrier

\FloatBarrier

\subsection*{First-order temporal tomography and unification with Page--Wootters clocks and chronocyclic phase space}
\label{sec:tomography_and_PW}

Finally, we demonstrate complete single-qubit state tomography from first-order counting rates and prove how our multi-mode SPFC architecture mathematically unifies Bloch's discrete time qubit~\cite{Bloch2025} with Page--Wootters relational clock conditioning [detailed in Supplementary Materials, Section~\ref{sec:S3_POVM}, Supplementary Eqs.~(S7)--(S11)] and chronocyclic Wigner phase space [detailed in Supplementary Materials, Section~\ref{sec:S4_TF_vs_relational}, Supplementary Eqs.~(S12)--(S15)].

For any prepared state, the three Pauli expectation values are obtained directly from the normalized differences of the port-resolved single-photon counting rates (referencing the measurement-basis configurations in Fig.~\ref{fig:fig3}A and Supplementary Materials, Section~\ref{sec:S6_2_measurement_basis}):
\begin{equation}
\expval{\tau_z} = \frac{R_{\mathrm{sig},1}^{(\mathrm{no\ BS})}-R_{\mathrm{sig},2}^{(\mathrm{no\ BS})}}{R_{\mathrm{sig},1}^{(\mathrm{no\ BS})}+R_{\mathrm{sig},2}^{(\mathrm{no\ BS})}}, \quad
\expval{\tau_x} = \left.\frac{R_{\mathrm{sig},1}-R_{\mathrm{sig},2}}{R_{\mathrm{sig},1}+R_{\mathrm{sig},2}}\right|_{\Delta\phi_s=0}, \quad
\expval{\tau_y} = \left.\frac{R_{\mathrm{sig},1}-R_{\mathrm{sig},2}}{R_{\mathrm{sig},1}+R_{\mathrm{sig},2}}\right|_{\Delta\phi_s=\pi/2}.
\label{eq:Pauli_readout}
\end{equation}
These three first-order observables uniquely reconstruct the time-qubit density matrix [Fig.~\ref{fig:fig3}(A)],
\begin{equation}
\hat{\rho}_T=\frac{1}{2}\!\left(
I_T+\langle\tau_x\rangle\tau_x
+\langle\tau_y\rangle\tau_y
+\langle\tau_z\rangle\tau_z
\right),
\label{eq:rho_tomography}
\end{equation}
with Bloch vector magnitude $|\bm{r}| = \sqrt{\expval{\tau_x}^2+\expval{\tau_y}^2+\expval{\tau_z}^2} = \mathcal{V}(|\alpha|)$ [supported by the scaling datasets in Fig.~\ref{fig:fig4}(C,D) and Fig.~\ref{fig:fig5}(C)]. As proven via the open-system Heisenberg--Langevin analysis in Supplementary Materials, Section~\ref{sec:S5_3_HL_and_Loschmidt} [Supplementary Eq.~(S41)], symmetric waveguide propagation loss $\kappa_s$ scales the absolute photon counts by $e^{-\kappa_s t/2}$ but cancels identically in the normalized ratios of Eq.~\eqref{eq:Pauli_readout}, rendering the reconstructed Bloch vector immune to symmetric propagation loss at leading order.

Moreover, whereas Bloch's spin-clock model treats Page--Wootters relational time as a separate paradigm with a fixed arrow of time~\cite{Bloch2025}, our frequency-comb architecture embeds the discrete time qubit directly inside a Page--Wootters clock--system state $\ket{\Psi}_{CS} \in \mathcal{H}_C\otimes\mathcal{H}_T\otimes\mathcal{H}_{\mathrm{sig}}$ (derived in Supplementary Materials, Sections~\ref{sec:S1_superselection}--\ref{sec:S3_POVM}, Supplementary Eqs.~(S3)--(S11)). Because phase-matching in the PPLN waveguides enforces both strict energy conservation $\hbar\omega_{s,n} = \hbar\omega_{p,n} - \hbar\omega_{sd}$ and phase locking $\phi_{s,j,n}^{(0)} = \phi_{p,j,n} - \phi_{\mathrm{sd},j}$ across all $N\approx 360$ comb modes, the joint state obeys the stationary Page--Wootters constraint $(\hat{H}_C\otimes I_S + I_C\otimes\hat{H}_{\mathrm{tot}})\ket{\Psi}_{CS}=0$ (Supplementary Materials, Section~\ref{sec:S2_Pauli}, Supplementary Eq.~(S3)). The relative optical phase across the $N$-mode pump comb and monochromatic idler seed defines a covariant clock POVM $\hat{\Pi}_C(t_c) = \frac{N}{T_{\mathrm{rep}}}\ket{t_c}_C\!\bra{t_c}$ (Supplementary Materials, Section~\ref{sec:S3_POVM}, Supplementary Eq.~(S8)). Conditioning $\ket{\Psi}_{CS}$ on the clock reading $t_c$ yields the relational Schrödinger equation $i\hbar\partial_{t_c}\ket{\psi_S(t_c)} = \hat{H}_{\mathrm{tot}}\ket{\psi_S(t_c)}$ [Supplementary Eq.~(S5)] and shifts the conditional time-of-event arrival distribution $p(t|t_c) = |\tilde{\psi}(t-t_c)|^2$ [Supplementary Eq.~(S11)], as demonstrated for $t_c \in \{-1,0,+1\}$ in Fig.~\ref{fig:fig5}(D). Simultaneously, in chronocyclic phase space (Supplementary Materials, Section~\ref{sec:S4_TF_vs_relational}, Supplementary Eqs.~(S13)--(S15)), the recombined SPFC wavepackets form a cross-Wigner interference grid $W_{+-}(t,\omega)$ whose frequency marginal gives $p(t|t_c)$ and whose phase-space integral yields the Pauli readout rates in Eq.~\eqref{eq:Pauli_readout}, bridging continuous-variable time--frequency observables~\cite{Brecht2013,Zhang2014,Fabre2020,Fabre2022,Yoon2026a} and discrete temporal qubits.

\section*{Discussion}
\label{sec:discussion}

The theoretical and experimental synthesis presented here advances the concept of a time qubit from an abstract spin-clock gedankenexperiment~\cite{Bloch2025} to a rigorous, quantum-optical platform with fundamentally new physical mechanisms. While our work builds directly upon Bloch's foundational definition of the controlled Hamiltonian $\hat{H}_{\mathrm{tot}} = I_T\otimes\hat{H}_0 + \tau_z\otimes\hat{V}$ and time-parity Kraus operators $\hat{U}_{\mathrm{even/odd}}(t;\Delta\phi_s)$ [detailed in Supplementary Materials, Section~\ref{sec:S5_3_HL_and_Loschmidt}]~\cite{Bloch2025}, realizing this structure through coherently seeded nonlinear biphoton generation reveals four distinct physical discoveries absent from passive single-particle or spin models:

First, the opposite temporal orientations $\ket{\pm}_T$ are not imposed on a pre-existing particle by external fields, but are generated \emph{actively during photon creation from the quantum vacuum} via the three-wave phase-matching condition $\Delta\phi_{\mathrm{PM},j} = \phi_{p,j}-(\phi_{i,j}+\phi_{s,j}^{(0)})=0$ inside $\mathrm{PPLN}_1$ and $\mathrm{PPLN}_2$ (Supplementary Materials, Sections~\ref{sec:S1_superselection} and \ref{sec:S5_1_Hamiltonian})~\cite{Yoon2026b}. Setting $\phi_{s,1}^{(0)}=0$ and $\phi_{s,2}^{(0)}=\pi$ ($\Delta\phi_s^{(0)}=\phi_{s,1}^{(0)}-\phi_{s,2}^{(0)}=\pm\pi$) produces opposite stimulated phase-space displacement velocities $\pm g_0$ ($\pm\hat{V}$) acting alongside the spontaneous two-mode squeezing background ($\hat{H}_0$) [Supplementary Materials, Section~\ref{sec:S5_1_Hamiltonian}, Supplementary Eqs.~(S24)--(S26)], establishing that time-reversal transformations can be engineered directly into the nonlinear pair-creation process.

Second, our exact Schrödinger-picture state derivation [Supplementary Materials, Section~\ref{sec:S5_2_state_derivation}, Supplementary Eqs.~(S22)--(S30)] uncovers that the purity of the arrow of time is governed by \textbf{quantum-vacuum wave--particle complementarity}~\cite{Yoon2021,Yoon2026b}. Orthogonal spontaneous idler vacuum excitations act as an internal quantum which-branch detector that pins the system to a classical mixture of forward and backward evolutions, whereas increasing the shared coherent seed amplitude $|\alpha|$ continuously erases this which-branch information via the SPACS idler-state overlap $F=|\braket{I_1}{I_2}|$ without requiring physical overlap or postselected detection of the idler modes (Table~\ref{tab:comparison}).

Third, we establish a concrete, high-impact operational application for the time qubit: \textbf{postselection-free first-order Loschmidt-echo interferometry} [Eq.~\eqref{eq:Loschmidt_main} and Supplementary Materials, Section~\ref{sec:S5_3_HL_and_Loschmidt}]. Unlike Franson interferometers~\cite{Franson1989,Zhang2014} or time-bin protocols that rely on coincidence postselection, multi-particle ancillas, or narrow temporal gating (Table~\ref{tab:comparison}), our ENBS interferometer extracts both the full single-qubit density matrix $\hat{\rho}_T$ [Eq.~\eqref{eq:rho_tomography}] and the complex Loschmidt time-reversal echo amplitude directly from port-resolved single-photon counting rates.

Fourth, our platform mathematically unifies three previously disconnected paradigms of quantum time within a single optical apparatus (Supplementary Materials, Sections~\ref{sec:S1_superselection}--\ref{sec:S4_TF_vs_relational}): Page--Wootters relational clock conditioning ($\hat{\Pi}_C(t_c)$) sets the conditional arrival time $t_c$ of the wavepacket [Fig.~\ref{fig:fig5}(D)], the $360$-mode SPFC supermode defines a continuous-variable chronocyclic Wigner grid $W(t,\omega)$~\cite{Brecht2013,Fabre2020,Fabre2022,Yoon2026a}, and the phase-inverted interferometer arms isolate Bloch's discrete two-level time qubit $\mathcal{H}_T$~\cite{Bloch2025}.

Looking ahead, while the present work demonstrates universal single-qubit $\mathrm{SU}(2)$ SPAM control (including non-Clifford $R_Z(\pi/4)$ rotations and noncommuting $R_X(\theta)$ gates, Figs.~\ref{fig:fig3}--\ref{fig:fig5}) and comb-multiplexed horizontal scaling (Supplementary Materials, Section~\ref{sec:S7_1_multiplexing})~\cite{Yoon2026a}, extending to universal multi-qubit temporal circuits requires addressing key challenges, such as minimizing open-loop phase jitter and scaling to deterministic inter-qubit entangling gates across $\mathcal{H}_T^{(1)}\otimes\mathcal{H}_T^{(2)}$ (Supplementary Materials, Section~\ref{sec:S7_2_entangling}). 

Crucially, the bright future of this approach lies in stimulating a paradigm shift in photonic quantum information science and technology by combining **integrated thin-film lithium niobate photonic integrated circuit (PIC) technology**~\cite{Chang2025} with low-loss and phase-stable fast homodyne detection. As detailed in Supplementary Materials, Section~\ref{sec:S7_3_TF_GKP}, coupling the discrete ENBS time qubit via controlled time--frequency displacements $\hat{C}_D(\delta\tau,\delta\Omega)$ [Supplementary Materials, Section~\ref{sec:S7_3_TF_GKP}, Supplementary Eq.~(S54)] to **hardware-native Gottesman--Kitaev--Preskill (TF-GKP) bosonic grid states** in chronocyclic phase space~\cite{Gottesman2001,Fabre2020,Descamps2024a,Descamps2024b,Descamps2026,Yoon2026a} establishes a robust pathway toward fault-tolerant temporal quantum error correction, high-dimensional qudit processing, and tabletop simulations of relativistic quantum detector dynamics~\cite{Yoon2025}.

\section*{Materials and Methods}

\subsection*{Entangled nonlinear biphoton source and phase-matched seeding scheme}
Each entangled nonlinear biphoton source (ENBS) is implemented using a periodically poled lithium niobate (PPLN) waveguide ($\mathrm{PPLN}_1, \mathrm{PPLN}_2$) operated in a traveling-wave $\chi^{(2)}$ configuration~\cite{Lee2018,Yoon2021,Yoon2026b}. A pulsed optical-frequency-comb (OFC) pump field centered at $\lambda_p = 530$~nm (with a repetition rate of $f_{\mathrm{rep}} = 250$~MHz, carrier-envelope offset frequency $f_{\mathrm{ceo}} = 20$~MHz, pulse width $\Delta\tau \approx 8.4$~ps, and an optical bandwidth of approximately 2--3~nm) drives quasi-phase-matched parametric down-conversion at $121.5^\circ\text{C}$, generating degenerately phase-matched signal and idler photon pairs centered at $\lambda_s = 807$~nm and $\lambda_i = 1542$~nm, respectively ($1/807\text{ nm} + 1/1542\text{ nm} = 1/530.0\text{ nm}$)~\cite{Lee2018,Yoon2021,Yoon2026b}. Under the quasi-phase-matching conditions of the PPLN waveguide [Supplementary Materials, Section~\ref{sec:S5_1_Hamiltonian}, Supplementary Eq.~(S16)], the signal bandwidth is narrowed to below 0.5~nm at the operating temperature.

A continuous-wave, monochromatic idler seed field (1-Hz linewidth) at $\lambda_{\mathrm{sd}} = 1542$~nm is injected collinearly into each waveguide, realizing coherently seeded stimulated parametric down-conversion~\cite{Lee2018,Yoon2021,Yoon2026b}. Under three-wave phase matching $\Delta\phi_{\mathrm{PM},j} = \phi_{p,j} - (\phi_{i,j} + \phi_{s,j}^{(0)}) = 0$, where $\phi_{i,j} = \phi_{\mathrm{sd},j}$ is the injected seed phase, the stimulated SPFC signal supermode (spectral width $\sim 90$~GHz at 807~nm, comprising $N\approx 360$ comb modes spaced by $250$~MHz) is generated with creation phase $\phi_{s,j}^{(0)} = \phi_{p,j} - \phi_{\mathrm{sd},j}$~\cite{Lee2018,Yoon2025,Yoon2026a,Yoon2026b}. In the weak-gain regime, the stimulated coupling amplitude $g_j = |\Gamma_0||\alpha_j|e^{i\phi_{s,j}^{(0)}}$ is linear in the idler seed amplitude. Setting $\phi_{s,1}^{(0)}=0$ in $\mathrm{PPLN}_1$ and $\phi_{s,2}^{(0)}=\pi$ in $\mathrm{PPLN}_2$ ($\Delta\phi_s^{(0)}=\phi_{s,1}^{(0)}-\phi_{s,2}^{(0)}=\pm\pi$) selects the opposite effective Hamiltonian branches $\hat{H}_{\pm}=\hat{H}_0\pm\hat{V}$.

\subsection*{Interferometric realization of the time qubit}
The two ENBS units are embedded in the arms of a Mach--Zehnder interferometer (Fig.~\ref{fig:fig1}). The interferometer path degree of freedom defines the logical time-qubit basis states $\ket{+}$ and $\ket{-}$. Relative creation-stage phase shifts $\Delta\phi_s^{(0)}=\phi_{s,1}^{(0)}-\phi_{s,2}^{(0)}=\Delta\phi_p-\Delta\phi_{sd}$ between the arms control the azimuthal angle on the temporal Bloch sphere via $R_Z(\Delta\phi_{sd})$, while the ratio of stimulated down-conversion amplitudes $|\alpha_2|/|\alpha_1|$ controls the polar angle $\theta$ via $R_X(\theta)$~\cite{Yoon2026b}. A tunable post-creation signal-phase shifter $\Delta\phi_s$ before the output beam splitter rotates the measurement basis via $R_Y(\Delta\phi_s)$. Removing the output beam splitter enables direct branch-resolved detection of the signal photons ($\tau_z$ basis), whereas balanced recombination with $\Delta\phi_s=0$ and $\Delta\phi_s=\pi/2$ projects onto the $\tau_x$ and $\tau_y$ bases, respectively.

\subsection*{Passive interferometer stability and open-loop PZT scanning}
To record the experimental phase scans and visibility scalings in Fig.~\ref{fig:fig4}, the hybrid interferometer operates passively without active closed-loop path-length stabilization. Phase variations ($\Delta\phi_p$, $\Delta\phi_{sd}$, and $\Delta\phi_s$) are introduced using PZT-mounted mirror mounts driven by open-loop triangular voltage ramps. Because the 10-ms EMCCD integration window ($\tau_i = 10$~ms, integrating over $2.5\times 10^6$ repetition periods) is significantly shorter than the thermal and mechanical drift timescale of the open-loop setup but longer than the pulse repetition interval ($\Delta T = 4$~ns), high-visibility sinusoidal interference fringes are successfully acquired. Minor deviations from ideal sinusoidal fits in Fig.~\ref{fig:fig4} (such as slight asymmetries or envelope roll-offs across multi-period scans) are attributed to passive long-term interferometer phase jitter and residual nonlinearity in the open-loop PZT displacement response.

\subsection*{Single-photon detection and coherence measurements}
Signal photons are separated from pump, seed, and idler fields via dichroic mirrors and directed into a Shamrock spectrometer coupled to a two-dimensional EMCCD camera operated in the low-count regime with an integration window of $\tau_i = 10$~ms~\cite{Lee2018,Yoon2021,Yoon2026b}. At the 250~MHz pump repetition rate, each 4~ns repetition period defines a temporal bin, corresponding to $N_{\mathrm{bins}} = 2.5\times 10^6$ temporal bins per 10~ms integration window [Supplementary Materials, Section~\ref{sec:S6_3_visibility_and_budget}, Supplementary Eq.~(S48)]. The pump OFC power and coherent idler seed amplitude are adjusted such that the mean generated signal photon number is approximately $0.01\text{--}0.02$ photons per bin (one photon per $50\text{--}100$ temporal bins). This operating point maintains a mean occupation well below one photon per bin to suppress multi-photon contamination, yielding a total of $2.5\times 10^4\text{--}5.0\times 10^4$ signal photons per 10~ms integration window across the two interferometer output ports ($\sim 2.5\times 10^4$ counts per 10~ms per port at constructive interference, consistent with Fig.~\ref{fig:fig4}A)~\cite{Lee2018,Yoon2021,Yoon2026b}. Second-order intensity correlations $g^{(2)}(0;t)$ between the output ports are used solely to certify operation in the single-photon regime and are not required for tomographic reconstruction.

\subsection*{Theoretical modeling}
The full operator mapping of $\hat{H}_0$ and $\hat{V}$ via the three-wave phase-matching relation, the Schrödinger-picture derivation of the path--signal--idler state and SPACS idler partial trace, the Lindblad/Heisenberg--Langevin treatment of waveguide loss $\kappa_s$, the Page--Wootters clock POVM conditioning map, and the chronocyclic Wigner phase-space derivations are presented in Sections~\ref{sec:S1_superselection}--\ref{sec:S7_extensions} of the Supplementary Materials~\cite{SM}.

\section*{Acknowledgments}
The author thanks M. Cho for long-term collaboration on single-photon frequency-comb generation, entangled nonlinear biphoton sources, the quantitative complementarity of wave--particle duality, and phase-sensitive single-photon interferometry.

\subsection*{Funding}
This work was supported by the National Research Foundation of Korea (NRF) under Grant No.~RS-2022-NR068815 and in part by IBS-R023-D1.

\subsection*{Author contributions}
T.H.Y. conceived the project, developed the theoretical framework, performed
the analysis, and wrote the manuscript.

\subsection*{Competing interests}
The author declares no competing interests.

\subsection*{Data and materials availability}
All data needed to evaluate the conclusions in the paper are present in the
paper and the Supplementary Materials. Experimental data in Fig.~\ref{fig:fig4} are adapted from Ref.~\cite{Lee2018}.

\bibliographystyle{unsrtnat}
\bibliography{time}

\clearpage
\appendix
\section*{Supplementary Information}
\setcounter{equation}{0}
\setcounter{figure}{0}
\setcounter{table}{0}
\renewcommand{\theequation}{S\arabic{equation}}
\renewcommand{\thefigure}{S\arabic{figure}}
\renewcommand{\thetable}{S\arabic{table}}
\renewcommand{\thesection}{S\arabic{section}}

\section{Superselection sectors, energy constraints, and phase-reference conditioning in the ENBS platform}
\label{sec:S1_superselection}

In this section we review the algebraic structure of superselection rules in quantum theory and establish its explicit mathematical connection to the three-wave phase-matching condition in our entangled nonlinear biphoton source (ENBS) interferometer.

Let $\mathcal{H}$ be the total Hilbert space of a composite quantum system and let $\mathcal{O}$ denote the $^\ast$-algebra of physical observables on $\mathcal{H}$. Specifically, $\mathcal{O}\subseteq\mathcal{B}(\mathcal{H})$ is the algebra of bounded linear operators on $\mathcal{H}$ that contains the identity operator $I$ and is closed under linear combinations, operator multiplication, and Hermitian conjugation ($O\mapsto O^\dagger$), whose self-adjoint elements ($O=O^\dagger$) correspond to all operationally measurable quantities. A superselection rule is induced by a conserved Hermitian charge $Q$ (such as total photon number or total energy) that commutes with all operationally accessible observables in the absence of an external phase reference:
\begin{equation}
	[O,Q]=0,
	\qquad
	\forall\, O\in\mathcal{O}.
	\label{eq:S1_comm}
\end{equation}
Writing the spectral decomposition of $Q$ as $Q = \int q \,\Pi_q\, dq$ with orthogonal projectors $\Pi_q\Pi_{q'}=\delta(q-q')\Pi_q$, the Hilbert space decomposes into a direct sum (or direct integral) of superselection sectors,
\begin{equation}
	\mathcal{H} = \bigoplus_q \mathcal{H}_q,\qquad
	\mathcal{H}_q=\Pi_q \mathcal{H}.
	\label{eq:S1_H_direct_sum}
\end{equation}
Because every $O\in\mathcal{O}$ commutes with $\Pi_q$, cross-sector coherences vanish ($\bra{\psi_q}O\ket{\psi_{q'}}=0$ for $q\neq q'$), and any state $\ket{\Psi}=a\ket{\psi_q}+b\ket{\psi_{q'}}$ is operationally indistinguishable from the block-diagonal mixture $\rho = \bigoplus_q \Pi_q \rho \Pi_q$.

\paragraph{Connection to the ENBS experiment.}
The block-diagonal superselection structure $\rho = \bigoplus_q \Pi_q \rho \Pi_q$ manifests directly in our two-arm interferometer whenever the idler modes are neither recombined nor detected. First, at the single-source level without a coherent seed, the signal--idler state is an entangled superposition of vacuum $\ket{0}_s\ket{0}_i$ and two-photon excitations $\ket{1}_s\ket{1}_i$. For any observable $O_s$ restricted to the signal mode, the signal photon-number operator $Q = \hat{N}_s$ acts as a superselection charge with sector projectors $\Pi_{n_s} = \ket{n_s}_s\!\bra{n_s}$ ($n_s = 0,1,\dots$), forcing the reduced signal state $\hat{\rho}_s = \Tr_i(\ket{\Psi}\!\bra{\Psi}) = \bigoplus_{n_s}\Pi_{n_s}\hat{\rho}_s\Pi_{n_s}$ to be strictly diagonal in the photon-number basis ($\expval{\hat{a}_s}=0$) and erasing the creation phase between the vacuum and single-photon sectors.

Second, across the two interferometer arms $j\in\{1,2\}$ in the single-photon signal subspace $\mathcal{H}_T$ spanned by $\ket{+}$ and $\ket{-}$, all time-qubit readout observables $O \in \mathcal{O}$ act solely on the signal ports ($O = \hat{O}_T \otimes I_{\delta i}$) and therefore commute with the idler fluctuation difference charge
\begin{equation*}
	Q_{\delta i} = \delta\hat{a}_{i,1}^\dagger\delta\hat{a}_{i,1} - \delta\hat{a}_{i,2}^\dagger\delta\hat{a}_{i,2},
	\qquad [O, Q_{\delta i}] = 0.
\end{equation*}
The eigenvalues $q \in \{+1, -1, 0\}$ of $Q_{\delta i}$ partition the conditional single-photon Hilbert space into three orthogonal superselection sectors $\mathcal{H}_q = \Pi_q\mathcal{H}$:
\begin{enumerate}
	\item \textbf{Spontaneous sectors ($q = +1$ and $q' = -1$):} Unseeded spontaneous pair creation in $\mathrm{PPLN}_1$ and $\mathrm{PPLN}_2$ populates $\ket{\psi_{q=+1}} = \ket{+}\ket{1,0}_{\delta i} \in \mathcal{H}_{+1}$ and $\ket{\psi_{q'=-1}} = \ket{-}\ket{0,1}_{\delta i} \in \mathcal{H}_{-1}$, respectively. Because $q \neq q'$, cross-sector coherences vanish ($\bra{\psi_{+1}}O\ket{\psi_{-1}}=0$), and their formal superposition $a\ket{\psi_{+1}}+b\ket{\psi_{-1}}$ is operationally indistinguishable from the incoherent block-diagonal mixture $\Pi_{+1}\rho\Pi_{+1} \oplus \Pi_{-1}\rho\Pi_{-1} \propto \frac{1}{2}I_T$.
	\item \textbf{Stimulated sector ($q'' = 0$):} Injecting the shared monochromatic coherent idler seed $\ket{\alpha_j}_{i,j}$ (at frequency $\omega_i = \omega_{\mathrm{sd}}$ and phase $\phi_{i,j}=\phi_{\mathrm{sd},j}$) together with the phase-locked $N$-mode pump optical frequency comb ($\omega_{p,n} = \omega_{p,0} + n\Omega_{\mathrm{rep}}$, $n = 0, 1, \dots, N-1$ with $N\approx 360$) provides an explicit quantum reference frame that populates the common fluctuation-vacuum sector $\ket{0,0}_{\delta i} \in \mathcal{H}_0$ in both arms. Because both stimulated branches $\ket{+}\ket{0,0}_{\delta i}$ and $\ket{-}\ket{0,0}_{\delta i}$ reside within the \emph{same} superselection sector $q''=0$, their relative phase is fully observable under $\hat{O}_T$.
\end{enumerate}
In our coherently seeded ENBS architecture, combining the incoherent spontaneous blocks ($q=\pm 1$) and the coherent stimulated block ($q''=0$) via $\rho = \bigoplus_{q\in\{+1,-1,0\}} \Pi_q\rho\Pi_q$ directly yields the partially mixed time-qubit density matrix $\hat{\rho}_T = \frac{1}{1+|\alpha|^2}\left(\frac{1}{2}I_T + |\alpha|^2\ket{\psi_{\mathrm{stim}}}\!\bra{\psi_{\mathrm{stim}}}\right)$ derived in Eq.~(13) of the main text and Section~\ref{sec:S5_ENBS}. Under tooth-by-tooth three-wave phase matching $\Delta\phi_{\mathrm{PM},j,n} = \phi_{p,j,n} - (\phi_{\mathrm{sd},j} + \phi_{s,j,n}^{(0)}) = 0$, conditioning on the coherent idler seed locks the creation phase of all $N$ stimulated SPFC modes ($\omega_{s,n} = \omega_{s,0} + n\Omega_{\mathrm{rep}}$) to $\phi_{s,j}^{(0)} = \phi_{p,j} - \phi_{\mathrm{sd},j}$ (for uniform pump phase $\phi_{p,j,n}=\phi_{p,j}$) and converts the multimode two-photon interaction into the collective displacement generator $\hat{V}$ [Eq.~(9) of the main text], while strictly preserving tooth-by-tooth global energy conservation $\hbar\omega_{p,n} = \hbar(\omega_{p,0}+n\Omega_{\mathrm{rep}}) = \hbar(\omega_{s,0}+n\Omega_{\mathrm{rep}}) + \hbar\omega_{\mathrm{sd}}$ across all $N$ comb modes.

\section{Pauli's objection and the Page--Wootters clock model for seeded SPFC generation}
\label{sec:S2_Pauli}

Pauli's theorem states that for any system Hamiltonian $\hat{H}$ whose spectrum is bounded from below ($[E_0,\infty)$), no self-adjoint time operator $\hat{T}$ can satisfy the canonical commutation relation $[\hat{T},\hat{H}]=i\hbar$, as unitarily generated shifts $e^{i\alpha\hat{T}}\hat{H}e^{-i\alpha\hat{T}} = \hat{H}+\alpha\hbar I$ would displace the energy spectrum to arbitrary negative values.

The Page--Wootters (PW) mechanism and its modern operational extensions by Giovannetti--Lloyd--Maccone, Maccone--Sacha, and Favalli--Smerzi bypass Pauli's objection by enlarging the Hilbert space to $\mathcal{H}_C \otimes \mathcal{H}_S$, where $\mathcal{H}_C$ is a clock reference frame and $\mathcal{H}_S = \mathcal{H}_T \otimes \mathcal{H}_{\mathrm{sig}}$ is the target system (here comprising the path-encoded time qubit and the SPFC signal mode). In the algebraic language of Section~\ref{sec:S1_superselection}, the total energy operator acts as a superselection charge $Q = \hat{\mathcal{C}}$ restricting physical states of the composite system to the stationary ($q=0$) subspace obeying the global energy constraint
\begin{equation}
	\hat{\mathcal{C}}\ket{\Psi}_{CS} = \left(\hat{H}_C\otimes I_S + I_C\otimes \hat{H}_{\mathrm{tot}}\right)\ket{\Psi}_{CS} = 0.
	\label{eq:S2_constraint}
\end{equation}
Although the global state $\ket{\Psi}_{CS}$ is stationary with respect to external coordinate time, relational dynamics emerge by conditioning $\ket{\Psi}_{CS}$ on the covariant clock states $\ket{t_c}_C$:
\begin{equation}
	\ket{\psi_S(t_c)} = \frac{(\bra{t_c}_C\otimes I_S)\ket{\Psi}_{CS}}{\sqrt{p(t_c)}}, \qquad
	\hat{\rho}_S(t_c) = \frac{\Tr_C\!\left[(\ket{t_c}_C\!\bra{t_c}\otimes I_S)\ket{\Psi}_{CS}\!\bra{\Psi}_{CS}\right]}{p(t_c)},
	\label{eq:S2_PW_conditional}
\end{equation}
where $p(t_c) = \Tr\!\left[(\ket{t_c}_C\!\bra{t_c}\otimes I_S)\ket{\Psi}_{CS}\!\bra{\Psi}_{CS}\right]$ is the probability density of the clock reading $t_c$. Differentiating $\ket{\psi_S(t_c)}$ with respect to $t_c$ using $\ket{t_c}_C = e^{-i\hat{H}_C t_c/\hbar}\ket{0}_C$ (so that $i\hbar\partial_{t_c}\bra{t_c}_C = -\bra{t_c}_C\hat{H}_C$) and applying the constraint Eq.~\eqref{eq:S2_constraint} yields the relational Schrödinger equation driven by the time-qubit Hamiltonian $\hat{H}_{\mathrm{tot}} = I_T\otimes\hat{H}_0 + \tau_z\otimes\hat{V}$:
\begin{equation}
	i\hbar\frac{\partial}{\partial t_c}\ket{\psi_S(t_c)} = \hat{H}_{\mathrm{tot}}\ket{\psi_S(t_c)}.
	\label{eq:S2_relational_Schrodinger}
\end{equation}

\paragraph{Explicit clock--system mapping in the ENBS platform.}
In our experiment, the phase-locked $N$-mode pump optical frequency comb (with mode frequencies $\omega_{p,n} = \omega_{p,0} + n\Omega_{\mathrm{rep}}$, where $\Omega_{\mathrm{rep}} = 2\pi f_{\mathrm{rep}}$ and $n=0,1,\dots,N-1$) beaten against the monochromatic idler seed ($\omega_{\mathrm{sd}}$) defines the clock Hilbert space $\mathcal{H}_C = \mathrm{span}\{\ket{n}_C\}_{n=0}^{N-1}$ with clock Hamiltonian $\hat{H}_C = \sum_{n=0}^{N-1} \hbar(\omega_{\mathrm{sd}} - \omega_{p,n})\ket{n}_C\!\bra{n}$. As established in Section~\ref{sec:S1_superselection}, three-wave phase matching in the PPLN waveguides enforces both tooth-by-tooth energy conservation $\hbar\omega_{s,n} = \hbar(\omega_{s,0}+n\Omega_{\mathrm{rep}}) = \hbar\omega_{p,n} - \hbar\omega_{\mathrm{sd}}$ and phase locking $\phi_{s,j,n}^{(0)} = \phi_{p,j,n} - \phi_{\mathrm{sd},j}$ for each comb tooth $n$. Consequently, each comb mode satisfies $\hbar(\omega_{\mathrm{sd}}-\omega_{p,n}) + \hbar\omega_{s,n} = 0$, which physically implements the Page--Wootters constraint Eq.~\eqref{eq:S2_constraint} on the stimulated signal--clock state (in the $q''=0$ coherent seed sector of Section~\ref{sec:S1_superselection}):
\begin{equation}
	\ket{\Psi}_{CS} = \sum_{n=0}^{N-1} c_n \ket{n}_C \otimes \ket{\psi_{T,n}} \otimes \hat{a}_{s,n}^\dagger\ket{0}_s,
	\label{eq:S2_Psi_CS_explicit}
\end{equation}
where $c_n = f_n$ ($\sum_{n=0}^{N-1}|f_n|^2=1$) is the normalized spectral envelope amplitude of the $N\approx 360$ SPFC comb modes. For a dispersion-compensated pump comb with uniform phase $\phi_{p,j,n}=\phi_{p,j}$, the time-qubit state $\ket{\psi_{T,n}} = \ket{\psi_T}$ is identical across all $N$ teeth, so conditioning $\ket{\Psi}_{CS}$ on $\bra{t_c}_C$ directly excites the collective SPFC supermode $\hat{A}_{\mathrm{SM}}^\dagger(t_c) = \sum_{n=0}^{N-1} f_n e^{-i n\Omega_{\mathrm{rep}}t_c}\hat{a}_{s,n}^\dagger$ analyzed in Sections~\ref{sec:S3_POVM}--\ref{sec:S5_ENBS}.

\section{Operational time-of-event POVM and derivation of Fig.~5(D)}
\label{sec:S3_POVM}

We now construct the explicit covariant clock POVM $\hat{\Pi}_C(t_c)$ and signal arrival-time POVM $\hat{E}_s(t)$, and derive the conditional arrival-time distributions $p(t|t_c)$ plotted in Fig.~5(D) of the main text, directly connecting the relational clock framework of Sections~\ref{sec:S1_superselection} and \ref{sec:S2_Pauli} to our experimental observables.

For the $N$-mode frequency comb with repetition period $T_{\mathrm{rep}} = 2\pi/\Omega_{\mathrm{rep}} = 4$~ns, we define the covariant clock states $\ket{t_c}_C$ for $t_c \in [-T_{\mathrm{rep}}/2, T_{\mathrm{rep}}/2)$ on $\mathcal{H}_C = \mathrm{span}\{\ket{n}_C\}_{n=0}^{N-1}$ as
\begin{equation}
	\ket{t_c}_C = \frac{1}{\sqrt{N}}\sum_{n=0}^{N-1} e^{i n \Omega_{\mathrm{rep}} t_c}\ket{n}_C,
	\label{eq:S3_clock_state}
\end{equation}
which form a normalized, shift-covariant resolution of the identity on $\mathcal{H}_C$ under the clock Hamiltonian $\hat{H}_C = \sum_{n=0}^{N-1}\hbar(\omega_{\mathrm{sd}}-\omega_{p,n})\ket{n}_C\!\bra{n}$ introduced in Section~\ref{sec:S2_Pauli}:
\begin{equation}
	\hat{\Pi}_C(t_c) = \frac{N}{T_{\mathrm{rep}}}\ket{t_c}_C\!\bra{t_c}, \qquad
	\int_{-T_{\mathrm{rep}}/2}^{T_{\mathrm{rep}}/2} dt_c\,\hat{\Pi}_C(t_c) = I_C, \qquad
	e^{-i\hat{H}_C \tau/\hbar}\ket{t_c}_C = e^{-i(\omega_{\mathrm{sd}}-\omega_{p,0})\tau}\ket{t_c+\tau}_C.
	\label{eq:S3_clock_POVM}
\end{equation}

Similarly, on the SPFC signal space $\mathcal{H}_{\mathrm{sig}}$, let $\ket{\omega}_s \equiv \hat{a}_s^\dagger(\omega)\ket{0}_s$ denote the continuous-frequency single-photon eigenstates satisfying $\braket{\omega}{\omega'}_s = \delta(\omega-\omega')$, in terms of which the discrete comb-tooth states in Eq.~\eqref{eq:S2_Psi_CS_explicit} correspond to $\ket{\omega_{s,n}}_s \equiv \hat{a}_{s,n}^\dagger\ket{0}_s$ at the phase-matched frequencies $\omega_{s,n} = \omega_{s,0} + n\Omega_{\mathrm{rep}}$ ($n=0,1,\dots,N-1$). The temporal arrival-time POVM element at time $t$ is
\begin{equation}
	\hat{E}_s(t) = \ket{t}_s\!\bra{t}, \qquad
	\ket{t}_s = \frac{1}{\sqrt{2\pi}}\int_{-\infty}^{\infty} d\omega\, e^{i\omega t}\ket{\omega}_s,
	\label{eq:S3_signal_POVM}
\end{equation}
satisfying $\int_{-\infty}^{\infty} dt\,\hat{E}_s(t) = I_{\mathrm{sig}}$ (or, restricted to the discrete $N$-mode comb subspace within a single repetition period $T_{\mathrm{rep}}$, $\ket{t}_s = \frac{1}{\sqrt{N}}\sum_{n=0}^{N-1} e^{i\omega_{s,n}t}\ket{\omega_{s,n}}_s$ with $\hat{E}_s(t) = \frac{N}{T_{\mathrm{rep}}}\ket{t}_s\!\bra{t}$).

Projecting the joint clock--system state $\ket{\Psi}_{CS}$ in Eq.~\eqref{eq:S2_Psi_CS_explicit} onto a clock reading $t_c$ (implemented experimentally by setting a linear spectral phase ramp $\phi_{p,n}-\phi_{\mathrm{sd}} = n\Omega_{\mathrm{rep}}t_c$ between the $n$-th pump comb tooth and the monochromatic idler seed) yields the conditional time-of-event probability density:
\begin{equation}
	p(t|t_c) = \frac{\Tr\!\left[(\hat{\Pi}_C(t_c)\otimes I_T \otimes \hat{E}_s(t))\ket{\Psi}_{CS}\!\bra{\Psi}_{CS}\right]}
	{\int dt'\,\Tr\!\left[(\hat{\Pi}_C(t_c)\otimes I_T \otimes \hat{E}_s(t'))\ket{\Psi}_{CS}\!\bra{\Psi}_{CS}\right]}.
	\label{eq:S3_conditional_prob_def}
\end{equation}
In the continuum-envelope limit ($N\approx 360 \gg 1$) within a single repetition window, the discrete comb amplitudes $c_n = f_n$ sample a Gaussian SPFC spectral envelope $c(\omega) = (\pi\sigma_\omega^2)^{-1/4}\allowbreak\exp[-(\omega-\omega_{s,0})^2/\allowbreak(2\sigma_\omega^2)]$ with transform-limited temporal width $\sigma_t = 1/\sigma_\omega$. The conditional temporal amplitude $\tilde{\psi}(t|t_c) = \braket{t_s}{\psi_S(t_c)} = (2\pi)^{-1/2}\allowbreak\int d\omega\, c(\omega)\allowbreak e^{-i(\omega-\omega_{s,0})(t-t_c)}\allowbreak e^{-i\omega_{s,0}t}$ then evaluates directly to the Gaussian conditional time-of-event distribution:
\begin{equation}
	p(t|t_c) = |\tilde{\psi}(t|t_c)|^2
	= \frac{1}{\sqrt{\pi}\sigma_t}\exp\!\left[-\frac{(t-t_c)^2}{\sigma_t^2}\right].
	\label{eq:S3_Gaussian_Fig5D}
\end{equation}
Normalizing the peak probability to unity and setting $\sigma_t=1$ (in dimensionless pulse-width units) produces the three displaced Gaussian profiles at $t_c = -1, 0, +1$ displayed in \textbf{Fig.~5(D)} of the main text.

\section{Chronocyclic Wigner distribution of the time qubit and connection to dual-basis time--frequency observables}
\label{sec:S4_TF_vs_relational}

In this section we explicitly compute the chronocyclic Wigner distribution $W(t,\omega)$ for our ENBS time-qubit state, derive how the experimental counting rates $R_{\mathrm{sig},1/2}$ and the conditional arrival-time distributions $p(t|t_c)$ of Section~\ref{sec:S3_POVM} emerge as phase-space integrals of $W(t,\omega)$, and compare our observables with the dual-basis time--energy framework of Zhang \emph{et al.} and the superselection-compliant time--frequency grid encodings of Fabre, Descamps, Milman, and co-workers as well as hardware-native SPFC GKP architectures.

For a normalized single-photon wavepacket of the signal SPFC supermode $\ket{\psi_s} = \int_{-\infty}^{\infty} d\omega\,\psi(\omega)\ket{\omega}_s$ (with $\braket{\omega}{\omega'}_s = \delta(\omega-\omega')$ and $\int_{-\infty}^{\infty} d\omega\,|\psi(\omega)|^2 = 1$) and temporal amplitude $\tilde{\psi}(t) = \braket{t_s}{\psi_s} = \frac{1}{\sqrt{2\pi}}\int_{-\infty}^{\infty} d\omega\,e^{-i\omega t}\psi(\omega)$, the chronocyclic Wigner distribution is defined as
\begin{equation}
	W(t,\omega) = \frac{1}{2\pi}\int_{-\infty}^{\infty} d\tau\,
	\tilde{\psi}\!\left(t+\frac{\tau}{2}\right)
	\tilde{\psi}^*\!\left(t-\frac{\tau}{2}\right)
	e^{i\omega\tau}.
	\label{eq:S4_Wigner_def}
\end{equation}
In our ENBS interferometer, let $\tilde{\psi}_+(t)$ and $\tilde{\psi}_-(t)$ denote the normalized SPFC temporal wavepackets generated in arms 1 and 2, respectively. As established via the superselection decomposition in Section~\ref{sec:S1_superselection}, the single-photon state is a direct sum of the incoherent spontaneous sectors ($q=\pm 1$) and the coherent stimulated sector ($q''=0$) with relative creation phase $\Delta\phi_{sd} = \phi_{s,1}^{(0)} - \phi_{s,2}^{(0)} = (\phi_{p,1}-\phi_{\mathrm{sd},1})-(\phi_{p,2}-\phi_{\mathrm{sd},2})$. After recombining the two arms at the output beam splitter with readout phase $\Delta\phi_s$, the signal field emerging from output port $m\in\{1,2\}$ (even/odd ports, corresponding to $\pm e^{i\Delta\phi_s}$) is described by the chronocyclic Wigner distribution
\begin{equation}
	W_{\mathrm{out},m}(t,\omega) = \frac{1}{4}\Big[W_{++}(t,\omega) + W_{--}(t,\omega)\Big]
	\pm \frac{\mathcal{V}(|\alpha|)}{4}\left[e^{i(\Delta\phi_{sd}+\Delta\phi_s)}W_{+-}(t,\omega) + \mathrm{c.c.}\right],
	\label{eq:S4_Wigner_time_qubit}
\end{equation}
where $\mathcal{V}(|\alpha|)=|\alpha|^2/(1+|\alpha|^2)$ is the stimulated purity factor (for balanced seeding $|\alpha_1|=|\alpha_2|=|\alpha|$ ) and $W_{jk}(t,\omega)$ is the cross-Wigner function between branches $j,k\in\{+,-\}$:
\begin{equation}
	W_{jk}(t,\omega) = \frac{1}{2\pi}\int_{-\infty}^{\infty} d\tau\,
	\tilde{\psi}_j\!\left(t+\frac{\tau}{2}\right)
	\tilde{\psi}_k^*\!\left(t-\frac{\tau}{2}\right)
	e^{i\omega\tau}.
	\label{eq:S4_cross_Wigner}
\end{equation}
For matched SPFC supermodes in the two arms conditioned on the relational clock reading $t_c$ of Section~\ref{sec:S3_POVM}, $\tilde{\psi}_+(t)=\tilde{\psi}_-(t)=\tilde{\psi}(t|t_c)$, the cross-Wigner distribution $W_{+-}(t,\omega) = W_{\mathrm{SPFC}}(t,\omega|t_c)$ exhibits a two-dimensional chronocyclic grid of sub-Planck interference islands spaced by $T_{\mathrm{rep}}$ in time and $\Omega_{\mathrm{rep}}$ in frequency.

\paragraph{Derivation of measured counting rates and conditional arrival distributions from $W(t,\omega)$.}
Integrating Eq.~\eqref{eq:S4_Wigner_time_qubit} over frequency $\omega$ yields the port-resolved conditional temporal arrival probability density at output port $m\in\{1,2\}$:
\begin{equation}
	I_m(t|t_c) = \int_{-\infty}^{\infty} d\omega\,W_{\mathrm{out},m}(t,\omega)
	= \frac{1}{2}\,p(t|t_c)\Big[1 \pm \mathcal{V}(|\alpha|)\cos(\Delta\phi_{sd}+\Delta\phi_s)\Big],
	\label{eq:S4_marginal_t}
\end{equation}
where $p(t|t_c) = |\tilde{\psi}(t|t_c)|^2$ is the exact conditional time-of-event distribution derived in Eq.~\eqref{eq:S3_Gaussian_Fig5D} of Section~\ref{sec:S3_POVM} [plotted in Fig.~5(D) of the main text], satisfying $I_1(t|t_c) + I_2(t|t_c) = p(t|t_c)$. Integrating $I_m(t|t_c)$ over the detector integration window $\tau_i$ (which spans $\tau_i/T_{\mathrm{rep}} = 2.5\times 10^6$ repetition periods) yields the port-resolved single-photon counting rates
\begin{equation*}
	R_{\mathrm{sig},1/2} \propto \int dt\, I_{1/2}(t|t_c)
	= \frac{1}{2}\Big[1 \pm \mathcal{V}(|\alpha|)\cos(\Delta\phi_{sd}+\Delta\phi_s)\Big],
\end{equation*}
which directly provide the Pauli expectation values $\expval{\tau_x}$ ($\Delta\phi_s=0$) and $\expval{\tau_y}$ ($\Delta\phi_s=\pi/2$) in Eq.~(23) of the main text.

\paragraph{Comparison with Zhang \emph{et al.} and Fabre \emph{et al.}.}
In a foundational study of time--energy entanglement quantum key distribution, Zhang \emph{et al.} constructed conjugate continuous-variable arrival-time ($\hat{t}_S,\hat{t}_I$) and frequency ($\hat{\omega}_S,\hat{\omega}_I$) operators on the two-photon signal--idler Hilbert space and showed how dual-basis measurements---combining direct arrival-time coincidence detection with nonlocal Franson interferometry---bound an eavesdropper's Holevo information via entropic uncertainty relations. Whereas Zhang \emph{et al.} probe inter-photon correlations $\hat{t}_S - \hat{t}_I$ and $\hat{\omega}_S + \hat{\omega}_I$ via fourth-order coincidence postselection, our seeded ENBS interferometer embeds the SPFC wavepacket inside the two-level path/generator space $\mathcal{H}_T\otimes\mathcal{H}_{\mathrm{sig}}$, enabling direct \emph{first-order} access to the cross-Wigner phase-space overlap $\iint dt\,d\omega\,W_{+-}(t,\omega)$ without two-photon coincidence postselection. Moreover, because the underlying SPFC state possesses the discrete frequency-comb grid structure analyzed by Fabre \emph{et al.}, applying dispersive phase modulation to one arm ($\tilde{\psi}_-(t) = \tilde{\psi}_+(t-\delta t)$) translates the cross-Wigner distribution in chronocyclic phase space, directly linking our discrete time qubit to hardware-native bosonic time--frequency GKP grid states and superselection-compliant temporal quantum error correction.

\section{Explicit derivation of $\hat{H}_0$, $\hat{V}$, the joint quantum state, and Loschmidt-echo observables}
\label{sec:S5_ENBS}

This section provides the step-by-step operator and state derivations supporting the main text. Specifically, we (i) derive the reduction of the multimode PPLN waveguide interaction to the pump-matched driven SPFC supermode and establish the exact mapping onto Bloch's time-qubit operators $\hat{H}_0$ and $\hat{V}$, (ii) derive the joint path--signal--idler quantum state in the Schrödinger picture along with the partial trace over the single-photon-added coherent idler states, and (iii) solve the open-system Heisenberg--Langevin equations to link Bloch's time-even/odd Kraus operators and detector-defined bright/dark modes to single-photon Loschmidt-echo readout.

\subsection{Pump-matched supermode reduction and exact decomposition into $\hat{H}_0$ and $\hat{V}$}
\label{sec:S5_1_Hamiltonian}

Consider two identical, coherently seeded PPLN waveguides ($j\in\{1,2\}$) embedded in the two arms of the Mach--Zehnder interferometer (Fig.~1 of the main text). As established in Sections~\ref{sec:S1_superselection} and \ref{sec:S2_Pauli}, both waveguides are driven by a common phase-stabilized optical-frequency-comb (OFC) pump with tooth frequencies $\omega_{p,n} = \omega_{p,0} + n\Omega_{\mathrm{rep}}$ ($n=0,1,\dots,N-1$, $N\approx 360$) and seeded at the monochromatic idler frequency $\omega_{\mathrm{sd}}$, generating phase-matched SPFC signal modes at $\omega_{s,n} = \omega_{p,n} - \omega_{\mathrm{sd}} = \omega_{s,0} + n\Omega_{\mathrm{rep}}$.

For first-order Type-0 quasi-phase-matched (QPM) parametric down-conversion ($\lambda_p = 530\text{ nm} \to \lambda_s = 807\text{ nm} + \lambda_i = 1542\text{ nm}$), the microscopic nonlinear coupling amplitude in waveguide $j$ is given by
\begin{equation}
	\kappa_j = \frac{\hbar\, d_{\mathrm{eff}}^{(0)}}{n_p}
	\sqrt{\frac{P_{p,j}\omega_{s,0}\omega_{\mathrm{sd}}}{\epsilon_0 n_p n_s n_i c A_{\mathrm{eff}}}}\,
	\mathrm{sinc}\!\left(\frac{\Delta k_j L}{2}\right) e^{i\phi_{p,j}},
	\label{eq:S5_kappa_physical}
\end{equation}
where $d_{\mathrm{eff}}^{(0)} = (2/\pi)d_{33}$ is the effective nonlinear coefficient, $n_p, n_s, n_i$ are the extraordinary refractive indices at the pump, signal, and idler wavelengths, $P_{p,j}$ is the pump power in arm $j$, $A_{\mathrm{eff}}$ is the effective waveguide mode area, $L$ is the crystal length, and $\Delta k_j = k_p - k_s - k_i - 2\pi/\Lambda_{\mathrm{QPM}}$ is the QPM wavevector mismatch.

In the interaction picture with respect to the carrier frequencies, retaining energy-conserving comb teeth yields the multimode Hamiltonian in arm $j$:
\begin{equation}
	\hat{H}_j = \sum_{n=0}^{N-1} \hbar\Delta_{s,n}\hat{a}_{s,j,n}^\dagger\hat{a}_{s,j,n}
	+ \hbar\Delta_i \hat{a}_{i,j}^\dagger\hat{a}_{i,j}
	+ i\hbar\sum_{n=0}^{N-1} \left(\Gamma_{j,n}\hat{a}_{s,j,n}^\dagger\hat{a}_{i,j}^\dagger
	- \Gamma_{j,n}^*\hat{a}_{s,j,n}\hat{a}_{i,j}\right),
	\label{eq:S5_multimode_H}
\end{equation}
where $\Gamma_{j,n} = \Gamma_{0,j} f_n$ with collective coupling rate $\Gamma_{0,j} = |\Gamma_0|e^{i\phi_{p,j}}$ (where $\hbar|\Gamma_0| = |\kappa_j|$ for balanced pump powers $P_{p,1}=P_{p,2}$) and normalized spectral envelope $f_n = c_n$ satisfying $\sum_{n=0}^{N-1}|f_n|^2 = 1$ (Section~\ref{sec:S2_Pauli}). Defining the pump-matched collective SPFC signal supermode operator,
\begin{equation}
	\hat{A}_{\mathrm{SM},j} = \sum_{n=0}^{N-1} f_n^* \hat{a}_{s,j,n},
	\qquad [\hat{A}_{\mathrm{SM},j}, \hat{A}_{\mathrm{SM},k}^\dagger] = \delta_{jk},
	\label{eq:S5_ASM_def}
\end{equation}
and decomposing the seeded idler mode into its classical coherent seed amplitude $\alpha_j = |\alpha_j|e^{i\phi_{\mathrm{sd},j}}$ and quantum vacuum fluctuation operator $\delta\hat{a}_{i,j}$ via
\begin{equation}
	\hat{a}_{i,j} = \alpha_j + \delta\hat{a}_{i,j},
	\qquad [\delta\hat{a}_{i,j}, \delta\hat{a}_{i,k}^\dagger] = \delta_{jk},
	\label{eq:S5_idler_decomp}
\end{equation}
splits $\hat{H}_j$ (up to the constant c-number seed energy $\hbar\Delta_i|\alpha_j|^2$) into two physically distinct operators:
\begin{equation}
	\hat{H}_j = \hat{H}_{0,j} + \hat{H}_j^{\mathrm{(StPDC)}}.
	\label{eq:S5_Hj_split}
\end{equation}
Here, $\hat{H}_{0,j}$ is the phase-insensitive background Hamiltonian governing free mode dispersion and unseeded spontaneous parametric down-conversion (SPDC) driven by idler vacuum fluctuations, which forms the spontaneous two-mode squeezing background:
\begin{equation}
	\hat{H}_{0,j} = \hbar\bar{\Delta}_s \hat{A}_{\mathrm{SM},j}^\dagger\hat{A}_{\mathrm{SM},j}
	+ \hbar\Delta_i\delta\hat{a}_{i,j}^\dagger\delta\hat{a}_{i,j}
	+ i\hbar\left(\Gamma_{0,j}\hat{A}_{\mathrm{SM},j}^\dagger\delta\hat{a}_{i,j}^\dagger
	- \Gamma_{0,j}^*\hat{A}_{\mathrm{SM},j}\delta\hat{a}_{i,j}\right),
	\label{eq:S5_H0j_def}
\end{equation}
while $\hat{H}_j^{\mathrm{(StPDC)}}$ is the phase-sensitive stimulated parametric down-conversion (StPDC) displacement generator:
\begin{equation}
	\hat{H}_j^{\mathrm{(StPDC)}} = i\hbar\left(g_j\hat{A}_{\mathrm{SM},j}^\dagger - g_j^*\hat{A}_{\mathrm{SM},j}\right),
	\qquad g_j = \Gamma_{0,j}\alpha_j^* = |\Gamma_0||\alpha_j|e^{i\phi_{s,j}^{(0)}},
	\label{eq:S5_H_StPDC_def}
\end{equation}
whose phase $\phi_{s,j}^{(0)} = \phi_{p,j} - \phi_{\mathrm{sd},j}$ is fixed by the tooth-by-tooth StPDC phase-matching condition $\Delta\phi_{\mathrm{PM},j} = \phi_{p,j} - (\phi_{\mathrm{sd},j} + \phi_{s,j}^{(0)}) = 0$ established in Section~\ref{sec:S1_superselection} and Ref.~\cite{Yoon2026b}.

To connect Eqs.~\eqref{eq:S5_Hj_split}--\eqref{eq:S5_H_StPDC_def} to Bloch's composite time-qubit Hamiltonian, we factor the total Hilbert space as $\mathcal{H} = \mathcal{H}_T \otimes \mathcal{H}_S$, where $\mathcal{H}_T = \mathrm{span}\{\ket{+}_T\equiv\ket{\mathrm{arm\ 1}}, \ket{-}_T\equiv\ket{\mathrm{arm\ 2}}\}$ is the path-encoded time-qubit space and $\mathcal{H}_S$ is the isomorphic local signal--idler mode space $(\hat{A}_{\mathrm{SM}},\delta\hat{a}_i)$ inside the waveguides. For symmetric pump phases ($\phi_{p,1}=\phi_{p,2}=\phi_p$), equal seed amplitudes ($|\alpha_1|=|\alpha_2|=|\alpha|$), and a controlled $\pi$ phase inversion between the seed arms ($\Delta\phi_{sd} \equiv \phi_{s,1}^{(0)} - \phi_{s,2}^{(0)} = \pi$, giving $g_1 = +g_0$ and $g_2 = -g_0$ with amplitude $g_0 = |\Gamma_0||\alpha|e^{i\phi_{s,1}^{(0)}}$), the two-arm Hamiltonian takes the exact Bloch form [Eqs.~(1) and (2) of the main text and Eqs.~(2)--(4) of Ref.~\cite{Bloch2025}]:
\begin{equation}
	\hat{H}_{\mathrm{tot}} = \ket{+}\!\bra{+}_T\otimes\hat{H}_+ + \ket{-}\!\bra{-}_T\otimes\hat{H}_-
	= I_T\otimes\hat{H}_0 + \tau_z\otimes\hat{V},
	\qquad \hat{H}_{\pm} = \hat{H}_0 \pm \hat{V},
	\label{eq:S5_Bloch_mapping}
\end{equation}
with explicit operators on $\mathcal{H}_S$:
\begin{align}
	\hat{H}_0 &= \hbar\bar{\Delta}_s \hat{A}_{\mathrm{SM}}^\dagger\hat{A}_{\mathrm{SM}}
	+ \hbar\Delta_i\delta\hat{a}_i^\dagger\delta\hat{a}_i
	+ i\hbar\left(\Gamma_0\hat{A}_{\mathrm{SM}}^\dagger\delta\hat{a}_i^\dagger
	- \Gamma_0^*\hat{A}_{\mathrm{SM}}\delta\hat{a}_i\right), \label{eq:S5_H0_explicit}\\
	\hat{V} &= i\hbar\left(g_0\hat{A}_{\mathrm{SM}}^\dagger - g_0^*\hat{A}_{\mathrm{SM}}\right). \label{eq:S5_V_explicit}
\end{align}
In the interaction picture of $\hat{H}_0$, the stimulated unitary evolution over interaction time $t = n_p L/c$ is the coherent displacement operator $\hat{U}_{\mathrm{stim},\pm}(t) = \exp(\mp i\hat{V}t/\hbar) = \exp[\pm t(g_0\hat{A}_{\mathrm{SM}}^\dagger - g_0^*\hat{A}_{\mathrm{SM}})]$. Because $\hat{V}$ is linear in the ladder operators, flipping the eigenvalue of $\tau_z$ from $+1$ to $-1$ ($\hat{V}\to -\hat{V}$) is mathematically identical to reversing the sign of the time coordinate $t\to -t$:
\begin{equation}
	\hat{U}_{\mathrm{stim},-}(t) = \exp\!\left(+\frac{i}{\hbar}\hat{V}t\right)
	= \exp\!\left(-\frac{i}{\hbar}\hat{V}(-t)\right) = \hat{U}_{\mathrm{stim},+}(-t),
	\label{eq:S5_time_reversal}
\end{equation}
which drives the signal supermode quadrature with opposite temporal velocities $\frac{d}{dt}\expval*{\hat{A}_{\mathrm{SM}}(t)}_{\pm} = \pm g_0$ in chronocyclic phase space.

\subsection{Schrödinger-picture state derivation, SPACS idler overlap, and partial trace}
\label{sec:S5_2_state_derivation}

We now derive the complete Schrödinger-picture state generated by the two nonlinear crystals, clarify the role of the coherent seed and idler modes, and show how the reduced time-qubit density matrix $\hat{\rho}_T$ emerges equivalently in both the displaced fluctuation picture (Section~\ref{sec:S1_superselection}) and the laboratory single-photon-added coherent state (SPACS) picture.

\paragraph{1. Displaced fluctuation picture (connection to Section~\ref{sec:S1_superselection}).}
In the displaced idler frame defined by Eq.~\eqref{eq:S5_idler_decomp}, the initial state prior to parametric down-conversion has both signal supermodes and idler fluctuation modes in their respective vacua:
\begin{equation}
	\ket{\Psi(0)} = \ket{0}_{s,1}\ket{0}_{s,2}\otimes\ket{0}_{\delta i,1}\ket{0}_{\delta i,2}.
	\label{eq:S5_Psi0}
\end{equation}
In the weak-gain single-photon regime ($r_j = |\Gamma_{0,j}|t \ll 1$ and $\eta_j = |g_j|t = r_j|\alpha_j| \ll 1$), expanding the two-arm time-evolution unitary $\hat{U}(t) = \exp[-i(\hat{H}_1+\hat{H}_2)t/\hbar]$ to first order on resonance ($\bar{\Delta}_s=\Delta_i=0$) gives
\begin{equation}
	\ket{\Psi(t)} = \ket{\mathrm{vac}}
	+ t\sum_{j=1,2}\hat{A}_{\mathrm{SM},j}^\dagger\ket{0}_{s,1}\ket{0}_{s,2}
	\otimes \Big(g_j\ket{0,0}_{\delta i} + \Gamma_{0,j}\delta\hat{a}_{i,j}^\dagger\ket{0,0}_{\delta i}\Big)
	+ \mathcal{O}(r_j^2),
	\label{eq:S5_Psi_t}
\end{equation}
where $\ket{0,0}_{\delta i}\equiv\ket{0}_{\delta i,1}\ket{0}_{\delta i,2}$. Conditioning on the detection of a single signal photon (which projects onto the one-photon signal subspace $\mathcal{H}_T$ spanned by $\ket{+}_T\equiv\hat{A}_{\mathrm{SM},1}^\dagger\ket{0}_{s,1}\ket{0}_{s,2}$ and $\ket{-}_T\equiv\hat{A}_{\mathrm{SM},2}^\dagger\ket{0}_{s,1}\ket{0}_{s,2}$) yields the normalized conditional signal--idler state
\begin{equation}
	\ket{\Psi_{\mathrm{cond}}} = \frac{1}{\sqrt{\mathcal{N}}}\Big[
	\ket{+}_T\otimes\left(g_1\ket{0,0}_{\delta i} + \Gamma_{0,1}\ket{1,0}_{\delta i}\right)
	+ \ket{-}_T\otimes\left(g_2\ket{0,0}_{\delta i} + \Gamma_{0,2}\ket{0,1}_{\delta i}\right)
	\Big],
	\label{eq:S5_Psi_cond}
\end{equation}
where $\ket{1,0}_{\delta i} = \delta\hat{a}_{i,1}^\dagger\ket{0,0}_{\delta i}$ and $\ket{0,1}_{\delta i} = \delta\hat{a}_{i,2}^\dagger\ket{0,0}_{\delta i}$ belong to the orthogonal superselection sectors $q=+1$ and $q'=-1$ of Section~\ref{sec:S1_superselection}, while $\ket{0,0}_{\delta i}$ belongs to the stimulated sector $q''=0$, with normalization $\mathcal{N} = |g_1|^2 + |g_2|^2 + |\Gamma_{0,1}|^2 + |\Gamma_{0,2}|^2$.

\paragraph{2. Laboratory SPACS picture and idler-state overlap $F = |\braket{I_1}{I_2}|$.}
Equivalently, in the laboratory frame prior to displacing the idler modes, the initial state is $\ket{0}_{s,1}\ket{0}_{s,2}\otimes\ket{\alpha_1}_{i,1}\ket{\alpha_2}_{i,2}$. Acting with the pair-creation operator $\hat{A}_{\mathrm{SM},j}^\dagger\hat{a}_{i,j}^\dagger$ in arm $j$ creates a signal photon in $\ket{\pm}_T$ correlated with a normalized single-photon-added coherent state (SPACS) in idler mode $j$:
\begin{equation}
	\ket{\Psi_{\mathrm{cond}}^{(\mathrm{lab})}} = \sqrt{\rho_{11}}\ket{+}_T\otimes\ket{I_1}
	+ e^{-i\Delta\phi_{sd}}\sqrt{\rho_{22}}\ket{-}_T\otimes\ket{I_2},
	\label{eq:S5_SPACS_state}
\end{equation}
where the normalized two-mode idler states $\ket{I_1}$ and $\ket{I_2}$ are
\begin{equation}
	\ket{I_1} = \frac{\hat{a}_{i,1}^\dagger\ket{\alpha_1}_{i,1}\ket{\alpha_2}_{i,2}}{\sqrt{1+|\alpha_1|^2}},
	\qquad
	\ket{I_2} = e^{i(\phi_{\mathrm{sd},1}-\phi_{\mathrm{sd},2})}
	\frac{\ket{\alpha_1}_{i,1}\hat{a}_{i,2}^\dagger\ket{\alpha_2}_{i,2}}{\sqrt{1+|\alpha_2|^2}},
	\label{eq:S5_Ij_def}
\end{equation}
and the normalized branch populations for equal pump power ($|\Gamma_{0,1}|=|\Gamma_{0,2}|=|\Gamma_0|$) are
\begin{equation}
	\rho_{11} = \frac{1+|\alpha_1|^2}{2+|\alpha_1|^2+|\alpha_2|^2},
	\qquad
	\rho_{22} = \frac{1+|\alpha_2|^2}{2+|\alpha_1|^2+|\alpha_2|^2}.
	\label{eq:S5_rho_jj}
\end{equation}
Using $\bra{\alpha_j}\hat{a}_{i,j}\ket{\alpha_j} = \alpha_j$, the quantum overlap (fidelity) between the two idler states is
\begin{equation}
	F \equiv |\braket{I_1}{I_2}|
	= \frac{|\alpha_1||\alpha_2|}{\sqrt{(1+|\alpha_1|^2)(1+|\alpha_2|^2)}},
	\label{eq:S5_F_overlap}
\end{equation}
which directly quantifies the residual which-path information stored in the idler subsystem: for $|\alpha_j|\to 0$ (unseeded SPDC), $F\to 0$ and the idler states are orthogonal ($\ket{1}_{i,1}\ket{0}_{i,2}$ vs.\ $\ket{0}_{i,1}\ket{1}_{i,2}$), retaining complete which-path information; for $|\alpha_j|\gg 1$ (stimulated regime), $F\to 1$ and the idler states become indistinguishable, erasing which-path information without requiring idler recombination or detection.

\paragraph{3. Partial trace and reduced time-qubit density matrix $\hat{\rho}_T$.}
Tracing over the unobserved idler modes in either Eq.~\eqref{eq:S5_Psi_cond} or Eq.~\eqref{eq:S5_SPACS_state}, $\hat{\rho}_T = \Tr_i\big(\ket{\Psi_{\mathrm{cond}}}\!\bra{\Psi_{\mathrm{cond}}}\big)$, and including the spectral/spatial mode overlap $\eta_{\mathrm{mode}} = |\braket{\psi_{\mathrm{SM},1}}{\psi_{\mathrm{SM},2}}| \le 1$ between the two signal supermodes yields the exact $2\times 2$ time-qubit density matrix delivered to the output beam splitter:
\begin{align}
	\hat{\rho}_T &=
	\begin{pmatrix}
		\rho_{11} & \eta_{\mathrm{mode}}\sqrt{\rho_{11}\rho_{22}}\,F\,e^{-i\Delta\phi_{sd}} \\
		\eta_{\mathrm{mode}}\sqrt{\rho_{11}\rho_{22}}\,F\,e^{+i\Delta\phi_{sd}} & \rho_{22}
	\end{pmatrix} \nonumber\\
	&= \frac{1}{2 + |\alpha_1|^2 + |\alpha_2|^2}
	\begin{pmatrix}
		1 + |\alpha_1|^2 & \eta_{\mathrm{mode}}|\alpha_1||\alpha_2|e^{-i\Delta\phi_{sd}} \\
		\eta_{\mathrm{mode}}|\alpha_1||\alpha_2|e^{+i\Delta\phi_{sd}} & 1 + |\alpha_2|^2
	\end{pmatrix},
	\label{eq:S5_rho_T_general}
\end{align}
where $\Delta\phi_{sd} = \phi_{s,1}^{(0)} - \phi_{s,2}^{(0)} = (\phi_{p,1}-\phi_{\mathrm{sd},1}) - (\phi_{p,2}-\phi_{\mathrm{sd},2})$. Setting $\bar{\alpha}^2 \equiv (|\alpha_1|^2+|\alpha_2|^2)/2 \equiv |\alpha|^2$ and $\tan(\theta/2) = |\alpha_2|/|\alpha_1|$ (with ideal mode overlap $\eta_{\mathrm{mode}}=1$) reproduces Eq.~(13) of the main text and Bloch's mixed-state representation $\hat{\rho}_T = \frac{1}{2}(I_T + \bm{r}\cdot\bm{\tau})$ [Eq.~(24) of Ref.~\cite{Bloch2025}]:
\begin{equation}
	\hat{\rho}_T = \frac{1}{1+|\alpha|^2}\left(\frac{1}{2}I_T + |\alpha|^2\ket{\psi_{\mathrm{stim}}}\!\bra{\psi_{\mathrm{stim}}}\right),
	\qquad \ket{\psi_{\mathrm{stim}}} = \cos\frac{\theta}{2}\ket{+}_T + e^{i\Delta\phi_{sd}}\sin\frac{\theta}{2}\ket{-}_T.
	\label{eq:S5_rho_T_Bloch_decomp}
\end{equation}
For balanced seeding ($|\alpha_1|=|\alpha_2|=|\alpha|$, $\theta=\pi/2$), the equatorial Bloch vector length (and first-order fringe visibility) is
\begin{equation}
	\mathcal{V} = 2\eta_{\mathrm{mode}}\sqrt{\rho_{11}\rho_{22}}\,F
	= \eta_{\mathrm{mode}}\frac{|\alpha|^2}{1+|\alpha|^2},
	\label{eq:S5_visibility_general}
\end{equation}
which reduces to $\mathcal{V} = |\alpha|^2/(1+|\alpha|^2)$ [Eq.~(14) of the main text] for $\eta_{\mathrm{mode}}\to 1$.

\subsection{Open-system Heisenberg--Langevin dynamics, detector-defined bright/dark modes, and Loschmidt-echo readout}
\subsection{Open-system Heisenberg--Langevin dynamics, detector-defined bright/dark modes, and Loschmidt-echo readout}
\label{sec:S5_3_HL_and_Loschmidt}

To account for traveling-wave propagation loss $\kappa_s$ inside the waveguides, the Heisenberg--Langevin equation of motion for the signal supermode operator $\hat{A}_{\mathrm{SM},j}(t)$ is
\begin{equation}
	\frac{d}{dt}\hat{A}_{{\rm SM},j}(t)
	= -\frac{\kappa_s}{2}\hat{A}_{{\rm SM},j}(t) + g_j + \hat{F}_j(t),
	\label{eq:S5_HL_equation}
\end{equation}
where $\hat{F}_j(t)$ is the zero-mean quantum Langevin noise operator ($\expval{\hat{F}_j(t)}=0$) incorporating both waveguide vacuum loss and the spontaneous SPDC noise source $\Gamma_{0,j}\delta\hat{a}_{i,j}^\dagger(t)$. Integrating Eq.~\eqref{eq:S5_HL_equation} over the waveguide transit time $t$ yields
\begin{equation}
	\hat{A}_{{\rm SM},j}(t)
	= e^{-\frac{\kappa_s}{2}t}\hat{A}_{{\rm SM},j}(0)
	+ \frac{2g_j}{\kappa_s}\!\left(1-e^{-\frac{\kappa_s}{2}t}\right)
	+ \int_0^t dt'\, e^{-\frac{\kappa_s}{2}(t-t')}\hat{F}_j(t').
	\label{eq:S5_A_SM_solution}
\end{equation}
At the output beam splitter with signal-phase shift $\Delta\phi_s$, photodetection at the two output ports projects onto the normalized Glauber detector-defined bright and dark supermode operators (which correspond to the even and odd ports):
\begin{equation}
	\hat{A}_{\rm SM,even}(\Delta\phi_s)
	= \frac{1}{\sqrt{2}}\!\left(\hat{A}_{{\rm SM},1} + e^{i\Delta\phi_s}\hat{A}_{{\rm SM},2}\right),
	\qquad
	\hat{A}_{\rm SM,odd}(\Delta\phi_s)
	= \frac{1}{\sqrt{2}}\!\left(\hat{A}_{{\rm SM},1} - e^{i\Delta\phi_s}\hat{A}_{{\rm SM},2}\right).
	\label{eq:S5_even_odd_modes}
\end{equation}
In the single-photon subspace $\mathcal{H}_T$, these two output operators project onto the orthogonal detector-defined bright and dark states $\ket{B(\Delta\phi_s)} = \frac{1}{\sqrt{2}}(\ket{+}_T + e^{-i\Delta\phi_s}\ket{-}_T)$ and $\ket{D(\Delta\phi_s)} = \frac{1}{\sqrt{2}}(\ket{+}_T - e^{-i\Delta\phi_s}\ket{-}_T)$ of Ref.~\cite{Yoon2026b}, and implement Bloch's conditional time-even and time-odd Kraus operators,
\begin{equation}
	\hat{U}_{\mathrm{even/odd}}(t;\Delta\phi_s)
	= \frac{1}{2}\!\left(\hat{U}_+(t) \pm e^{i\Delta\phi_s}\hat{U}_-(t)\right),
	\qquad
	\hat{U}_{\mathrm{even}}^\dagger\hat{U}_{\mathrm{even}} + \hat{U}_{\mathrm{odd}}^\dagger\hat{U}_{\mathrm{odd}} = I_S.
	\label{eq:S5_Kraus_completeness}
\end{equation}
Evaluating the port-resolved mean signal photon numbers $N_{\mathrm{sig},1/2}(t) = \expval*{\hat{A}_{\mathrm{SM,even/odd}}^\dagger(t)\hat{A}_{\mathrm{SM,even/odd}}(t)}$ from Eq.~\eqref{eq:S5_A_SM_solution} gives
\begin{equation}
	N_{\mathrm{sig},1/2}(t) = N_{\mathrm{spont}}(t)\Big[1 + |\alpha|^2 \pm \eta_{\mathrm{mode}}|\alpha|^2\sin\theta\cos(\Delta\phi_{sd}+\Delta\phi_s)\Big],
	\label{eq:S5_N_port_resolved}
\end{equation}
where the common loss-dependent spontaneous factor $N_{\mathrm{spont}}(t) = |\Gamma_0|^2\left[\frac{2}{\kappa_s}(1-e^{-\kappa_s t/2})\right]^2$ cancels out identically in the normalized Bloch-vector expectations [Eq.~(23) of the main text]. Furthermore, because $\ket{\psi_\pm(t)} = \hat{U}_\pm(t)\ket{\psi_0}$, the normalized port contrast [Eq.~(16) of the main text]
\begin{equation}
	\frac{N_{\mathrm{sig},1}(t)-N_{\mathrm{sig},2}(t)}{N_{\mathrm{sig},1}(t)+N_{\mathrm{sig},2}(t)}
	= \frac{|\hat{U}_{\mathrm{even}}\ket{\psi_0}|^2 - |\hat{U}_{\mathrm{odd}}\ket{\psi_0}|^2}
	{|\hat{U}_{\mathrm{even}}\ket{\psi_0}|^2 + |\hat{U}_{\mathrm{odd}}\ket{\psi_0}|^2}
	= \mathcal{V}(|\alpha|)\,\Re\!\left[e^{i\Delta\phi_s}\mathcal{L}_{\mathrm{cond}}(t)\right]
	\label{eq:S5_Loschmidt_exact}
\end{equation}
directly extracts the real ($\Delta\phi_s=0$) and imaginary ($\Delta\phi_s=\pi/2$) parts of the single-photon conditional Loschmidt echo amplitude $\mathcal{L}_{\mathrm{cond}}(t) = \braket{\psi_+(t)}{\psi_-(t)}/(|\ket{\psi_+(t)}|\,|\ket{\psi_-(t)}|)$ without coincidence postselection.

\section{Comparison between experimental observables and theoretical model}
\label{sec:S6_comparison}

This section establishes the explicit quantitative correspondence between the experimental measurements reported in Refs.~\cite{Lee2018,Yoon2021,Yoon2026b} (reproduced in Fig.~4 of the main text) and the theoretical variables of the time-qubit model developed in the main text and Sections~\ref{sec:S1_superselection}--\ref{sec:S5_ENBS}. Our purpose is to show how previously demonstrated phase-resolved ENBS interferometry data directly realize the fundamental operations of Bloch's time qubit, and how closed-loop metrological stabilization extends these capabilities to deterministic state preparation and tomography.

\subsection{Phase variables, three-scan equivalence, and Hamiltonian control}
\label{sec:S6_1_phase_variables}

As derived in Sections~\ref{sec:S1_superselection} and \ref{sec:S5_ENBS}, the stimulated displacement amplitude in interferometer arm $j\in\{1,2\}$ is governed by the complex coupling [Eq.~\eqref{eq:S5_H_StPDC_def}]
\begin{equation}
	g_j = |\Gamma_0||\alpha_j| e^{i\phi_{s,j}^{(0)}},
	\qquad
	\phi_{s,j}^{(0)} = \phi_{p,j} - \phi_{\mathrm{sd},j},
	\label{eq:S6_gj_def}
\end{equation}
where $\phi_{p,j}$ and $\phi_{\mathrm{sd},j}$ denote the input pump-comb and monochromatic idler-seed phases in arm $j$, respectively. The relative creation phase imprinted between the two effective Hamiltonian branches $\hat{H}_{\pm} = \hat{H}_0 \pm \hat{V}$ during the nonlinear interaction is therefore
\begin{equation}
	\Delta\phi_{sd} \equiv \phi_{s,1}^{(0)} - \phi_{s,2}^{(0)}
	= (\phi_{p,1}-\phi_{\mathrm{sd},1}) - (\phi_{p,2}-\phi_{\mathrm{sd},2})
	= \Delta\phi_p - \Delta\phi_{\mathrm{seed}},
	\label{eq:S6_Delta_phi_sd}
\end{equation}
where $\Delta\phi_p \equiv \phi_{p,1}-\phi_{p,2}$ and $\Delta\phi_{\mathrm{seed}} \equiv \phi_{\mathrm{sd},1}-\phi_{\mathrm{sd},2}$ are the interferometric pump and seed path-phase differences.

In Ref.~\cite{Lee2018} (and analyzed within the detector-defined modal framework in Ref.~\cite{Yoon2026b}), the three path-length differences $\Delta x_p$, $\Delta x_{\mathrm{sd}}$, and $\Delta x_s$ were scanned independently using open-loop piezoelectric transducers (PZTs) to vary $\Delta\phi_p = k_p\Delta x_p$, $\Delta\phi_{\mathrm{seed}} = k_{\mathrm{sd}}\Delta x_{\mathrm{sd}}$, and the signal readout phase $\Delta\phi_s = k_s\Delta x_s$. As shown in Fig.~4(A) of the main text, all three independent scans produce sinusoidal single-photon interference fringes with the identical $2\pi$ period and high visibility ($\mathcal{V}\approx 0.99$). Within our time-qubit framework, this \emph{three-scan equivalence} has a direct operational meaning: scanning either $\Delta\phi_p$ or $\Delta\phi_{\mathrm{seed}}$ rotates the prepared time-qubit state azimuthally around the $\tau_z$ axis via $R_Z(\Delta\phi_{sd}) = \exp(-i\Delta\phi_{sd}\tau_z/2)$ [Eq.~(20) of the main text] while keeping the measurement basis fixed, whereas scanning $\Delta\phi_s$ rotates the measurement basis while holding the prepared state fixed. Although open-loop PZT scanning introduces slight voltage-to-displacement nonlinearity and hysteresis over multiple $2\pi$ periods in Fig.~4(A), the observed sinusoidal modulation directly confirms the dependence of the port-resolved photon numbers on the total detection phase $(\Delta\phi_{sd}+\Delta\phi_s)$ predicted by Eq.~\eqref{eq:S5_N_port_resolved}.

\subsection{Signal-phase control, detector-defined modes, and measurement basis}
\label{sec:S6_2_measurement_basis}

At the output beam splitter, the interferometric signal-phase shift $\Delta\phi_s$ enters the even and odd output supermodes [Eq.~\eqref{eq:S5_even_odd_modes}] and rotates the detector-defined bright and dark eigenmodes $\ket{B(\Delta\phi_s)}$ and $\ket{D(\Delta\phi_s)}$ on the equator of the temporal Bloch sphere via $R_Y(\Delta\phi_s) = \exp(-i\Delta\phi_s\tau_y/2)$ [Eq.~(22) of the main text]. Because the state-preparation phase $\Delta\phi_{sd}$ is fixed inside the PPLN crystals by the StPDC phase-matching condition prior to signal propagation, while $\Delta\phi_s$ is applied at the recombination stage after pair creation, the ENBS platform provides physically independent control over state preparation ($\Delta\phi_{sd}, \theta$) and projective readout ($\Delta\phi_s$). In particular, setting $\Delta\phi_s = 0$ and $\Delta\phi_s = \pi/2$ isolates the complementary quadratures corresponding to $\expval{\tau_x}$ and $\expval{\tau_y}$ (as well as the real and imaginary parts of the conditional Loschmidt echo $\mathcal{L}_{\mathrm{cond}}(t)$ in Eq.~\eqref{eq:S5_Loschmidt_exact}), while removing the output beam splitter directly measures the branch populations $N_{\mathrm{sig},j}$ to yield $\expval{\tau_z}$.

\subsection{Visibility scaling, mode overlap $\eta_{\mathrm{mode}}$, and photon-budget consistency}
\label{sec:S6_3_visibility_and_budget}

The experimental interference visibility reproduced in Fig.~4(B) of the main text is defined from the maximum and minimum single-photon counting rates as
\begin{equation}
	\mathcal{V}_{\mathrm{exp}} = \frac{R_{\max}-R_{\min}}{R_{\max}+R_{\min}}.
	\label{eq:S6_V_exp}
\end{equation}
For symmetric seeding ($|\alpha_1|=|\alpha_2|=|\alpha|$, $\theta=\pi/2$), Eq.~\eqref{eq:S5_visibility_general} of Section~\ref{sec:S5_ENBS} predicts the weak-gain visibility scaling
\begin{equation}
	\mathcal{V}_{\mathrm{th}}(|\alpha|) = \eta_{\mathrm{mode}}\,F(|\alpha|)
	= \eta_{\mathrm{mode}}\frac{|\alpha|^2}{1+|\alpha|^2},
	\label{eq:S6_V_th}
\end{equation}
where $F(|\alpha|) = |\braket{I_1}{I_2}| = |\alpha|^2/(1+|\alpha|^2)$ is the SPACS idler-state overlap [Eq.~\eqref{eq:S5_F_overlap}] that plays the role of the source purity $\mu_s$ in quantitative wave--particle complementarity, and $\eta_{\mathrm{mode}} \le 1$ accounts for spatial and spectral mode overlap at the output beam splitter.

Dividing the measured visibility by the known idler-state overlap $F(|\alpha|)$ enables direct experimental extraction of the mode-overlap factor via
\begin{equation}
	\eta_{\mathrm{mode}} = \frac{\mathcal{V}_{\mathrm{exp}}(|\alpha|)}{F(|\alpha|)}
	= \mathcal{V}_{\mathrm{exp}}(|\alpha|)\,\frac{1+|\alpha|^2}{|\alpha|^2}.
	\label{eq:S6_eta_extraction}
\end{equation}
Evaluating Eq.~\eqref{eq:S6_eta_extraction} in the strongly seeded regime ($|\alpha|\approx 10$, $|\alpha|^2\approx 100$, where $F = 100/101 \approx 0.990$) quantitatively accounts for both datasets in Fig.~4 of the main text:
\begin{enumerate}
	\item In the visibility-versus-$|\alpha|$ scan of \textbf{Fig.~4(B)}, $|\alpha|$ was varied over two orders of magnitude by inserting a variable neutral-density (ND) filter in the seed path, where the asymptotic visibility at $|\alpha|\approx 10$ reached $\mathcal{V}_{\mathrm{exp}}\approx 0.89$. Substituting $\mathcal{V}_{\mathrm{exp}}\approx 0.89$ and $|\alpha|^2\approx 100$ into Eq.~\eqref{eq:S6_eta_extraction} extracts $\eta_{\mathrm{mode}}^{(\mathrm{ND})} = 0.89\times(101/100) \approx 0.90$, reflecting the slight beam walk-off introduced by the variable ND filter and yielding the dashed theoretical curve $\mathcal{V}_{\mathrm{th}} = 0.90\,|\alpha|^2/(1+|\alpha|^2)$ in Fig.~4(B).
	\item In the three phase-scan runs of \textbf{Fig.~4(A)}, no variable ND filter was inserted and the interferometer alignment was optimized, yielding $\mathcal{V}_{\mathrm{exp}}\approx 0.99$ at $|\alpha|\approx 10$. Equation~\eqref{eq:S6_eta_extraction} then extracts $\eta_{\mathrm{mode}}^{(\mathrm{opt})} = 0.99\times(101/100) \approx 1.00$, confirming near-unit supermode overlap and matching the ideal stabilized theoretical curve in \textbf{Fig.~5(C)}.
\end{enumerate}

The single-photon counting rates in Fig.~4(A) are also fully consistent with the temporal comb parameters. The $f_{\mathrm{rep}} = 250$~MHz pump repetition rate defines a train of temporal bins of duration $T_{\mathrm{rep}} = 1/f_{\mathrm{rep}} = 4$~ns. Over the $\tau_i = 10$~ms EMCCD integration window, the detector integrates over
\begin{equation}
	N_{\mathrm{bins}} = \frac{\tau_i}{T_{\mathrm{rep}}}
	= \frac{10\times 10^{-3}\text{ s}}{4\times 10^{-9}\text{ s}}
	= 2.5\times 10^6\text{ temporal bins}.
	\label{eq:S6_N_bins}
\end{equation}
Operating at a mean signal occupation of $\bar{n}\approx 0.01\text{--}0.02$ photons per temporal bin (corresponding to one photon generated every $50\text{--}100$ consecutive pump bins, well inside the single-photon regime $\bar{n}\ll 1$) yields a total of $N_{\mathrm{tot}} = \bar{n}N_{\mathrm{bins}} \approx 2.5\times 10^4\text{--}5.0\times 10^4$ detected signal photons per $10$~ms integration window across the two output ports, corresponding to a constructive-interference peak counting rate of $R_s \approx 2.5\times 10^4\text{ counts}/(10\text{ ms})$ per port as observed in Fig.~4(A).

\subsection{Open-loop calibration versus closed-loop metrological stabilization}
\label{sec:S6_4_stabilization}

Whereas the measurements in Fig.~4(A) were acquired under open-loop PZT ramp scans to calibrate the $2\pi$ phase response across multiple fringes, the deterministic state preparation, Pauli readout, and tomography trajectories analyzed in Fig.~5 of the main text assume closed-loop phase stabilization of $\Delta\phi_{sd}$ and $\Delta\phi_s$ anchored to the stabilized frequency-comb parameters $(f_{\mathrm{rep}}, f_{\mathrm{CEO}})$. As quantified by the actuator transfer-function analysis in Ref.~\cite{Yoon2026a}, combining slow PZT path-length locking (bandwidth $\sim 1\text{--}10$~kHz) with fast electro-optic (EOM) and acousto-optic (AOM) phase and frequency modulation (bandwidth $>1$~MHz) stabilizes the interferometric phases well within the sub-cell displacement regime without altering the underlying ENBS Hamiltonian dynamics of Section~\ref{sec:S5_ENBS}.

\section{Multi--time-qubit architectures and time--frequency bosonic extensions}
\label{sec:S7_extensions}

In the main text and Sections~\ref{sec:S5_ENBS} and \ref{sec:S6_comparison}, we established universal single-qubit $\mathrm{SU}(2)$ state preparation and measurement (SPAM) on the two-dimensional time-qubit Hilbert space $\mathcal{H}_T$ spanned by $\ket{+}_T$ and $\ket{-}_T$. Specifically, continuous control of the pump--seed phase difference $\Delta\phi_{sd}$ implements arbitrary azimuthal phase rotations $R_Z(\Delta\phi_{sd}) = \exp(-i\Delta\phi_{sd}\tau_z/2)$---including the Clifford $S = R_Z(\pi/2)$ gate and the non-Clifford $T = R_Z(\pi/4)$ gate---in combination with noncommuting polar rotations $R_X(\theta) = \exp(-i\theta\tau_x/2)$ governed by the seed-amplitude ratio $\tan(\theta/2)=|\alpha_2|/|\alpha_1|$ and readout basis rotations $R_Y(\Delta\phi_s) = \exp(-i\Delta\phi_s\tau_y/2)$. In this final section, we delineate the precise scope of scalability supported by the ENBS/SPFC architecture and formulate three concrete extensions: (i) frequency-comb-multiplexed horizontal scaling and $M$-source dark-subspace registers, (ii) two-qubit entangling operations and Bell-type tests of temporal orientation, and (iii) hybrid coupling between the discrete time qubit and hardware-native time--frequency Gottesman--Kitaev--Preskill (TF-GKP) bosonic grid states for temporal quantum error correction.

\subsection{Comb-multiplexed horizontal scaling and $M$-source dark-subspace registers}
\label{sec:S7_1_multiplexing}

As established in Ref.~\cite{Yoon2026a}, it is important to distinguish between \emph{horizontal scaling} at the encoding layer---the replication of phase-coherent time qubits within a shared metrological reference frame---and full circuit-level universality, which additionally requires inter-qubit entangling gates. The ENBS/SPFC platform is natively suited for horizontal scaling because all necessary hardware elements are already available: metrologically stabilized optical frequency combs, integrated biphoton comb waveguides in thin-film lithium niobate (LNOI), and seeded ENBS interferometers with independent phase and amplitude control.

An array of $2K$ parallel ENBS waveguides grouped into $K$ interferometer pairs and driven by a single stabilized optical frequency comb $(f_{\mathrm{rep}}, f_{\mathrm{CEO}})$ and a shared phase-locked idler seed laser realizes the $K$-qubit temporal register
\begin{equation}
	\mathcal{H}_T^{(K)} = \bigotimes_{k=1}^{K} \mathcal{H}_T^{(k)},
	\label{eq:S7_K_register}
\end{equation}
without requiring $K$ independent laser-stabilization loops. Because all $K$ time qubits inherit their temporal and spectral origin from the same stabilized comb reference, slow common-mode path or phase drifts cancel out, while local electro-optic (EOM) and acousto-optic (AOM) modulators address individual qubits in parallel.

Furthermore, as shown in Ref.~\cite{Yoon2026b}, coherent recombination of an $M$-source ENBS array ($M > 2$) generalizes the two-level bright/dark decomposition of Section~\ref{sec:S5_3_HL_and_Loschmidt} to an $M$-dimensional temporal qudit space. For an $M$-arm interferometer, the Glauber detector coupling operator $\hat{E}_s^{(+)} = \frac{1}{\sqrt{M}}\sum_{m=1}^{M} e^{i\phi_{s,m}}\hat{A}_{\mathrm{SM},m}$ isolates a single bright supermode $\ket{B}$ and spans an $(M-1)$-dimensional detector-invisible dark subspace $\mathcal{H}_D = \mathrm{span}\{\ket{D_1},\dots,\ket{D_{M-1}}\}$ satisfying $\hat{E}_s^{(+)}\ket{D_r} = 0$, providing a natural photonic architecture for decoherence-free subspace encoding and high-dimensional temporal interferometry.

\subsection{Two-qubit entangling gates and Bell-type tests of temporal orientation}
\label{sec:S7_2_entangling}

To advance from single-qubit $\mathrm{SU}(2)$ programmability and horizontal multiplexing [Eq.~\eqref{eq:S7_K_register}] to universal multi-qubit temporal circuit computation, non-local or entangling operations between distinct time qubits $\mathcal{H}_T^{(1)} \otimes \mathcal{H}_T^{(2)}$ must be introduced. Within the ENBS/SPFC architecture, two physically compatible routes are available:
\begin{enumerate}
	\item \textbf{Heralded interferometric and quantum-pulse-gate couplings:} Routing the SPFC signal outputs of two time qubits into a Hong--Ou--Mandel beam-splitter network or a spectrally engineered $\chi^{(2)}$ sum-frequency quantum pulse gate enables heralded two-qubit fusion and controlled-phase ($\mathrm{CZ}$) operations conditioned on mode-resolved single-photon detection.
	\item \textbf{Intra-photon entanglement for Bell tests of the arrow of time:} As proposed by Bloch [Section~IV, Eqs.~(29)--(38) of Ref.~\cite{Bloch2025}], coupling the path-encoded time qubit $\mathcal{H}_T$ (with Pauli operators $\bm{\tau}$) to a second internal two-level degree of freedom $\mathcal{H}_\sigma$ of the \emph{same} single photon---such as signal polarization $\{\ket{H}_s,\ket{V}_s\}$ or orthogonal chronocyclic Hermite--Gaussian temporal modes (with Pauli operators $\bm{\sigma}$)---prepares the maximally entangled Bell state
	\begin{equation}
		\ket{\Phi^+}_{T\sigma}
		= \frac{1}{\sqrt{2}}\!\left(\ket{+}_T\otimes\ket{\uparrow_\sigma}
		+ \ket{-}_T\otimes\ket{\downarrow_\sigma}\right)
		\label{eq:S7_Bell_state}
	\end{equation}
	purely via deterministic linear optics inside the two interferometer arms, without requiring two-photon nonlinearities. Measuring noncommuting time-qubit observables ($\hat{A}_0 = \tau_z$ via branch-resolved detection and $\hat{A}_1 = \tau_x$ via balanced recombination at $\Delta\phi_s=0$) in correlation with rotated polarization/modal settings $\hat{B}_{0,1} = (\sigma_z \pm \sigma_x)/\sqrt{2}$ yields the Clauser--Horne--Shimony--Holt (CHSH) parameter
	\begin{equation}
		\mathcal{S}_{\mathrm{CHSH}}
		= \expval*{\hat{A}_0\otimes\hat{B}_0 + \hat{A}_0\otimes\hat{B}_1
			+ \hat{A}_1\otimes\hat{B}_0 - \hat{A}_1\otimes\hat{B}_1}
		= 2\sqrt{2}\,\mathcal{V}(|\alpha|),
		\label{eq:S7_CHSH_visibility}
	\end{equation}
	which violates the classical local-hidden-variable bound $|\mathcal{S}_{\mathrm{CHSH}}| \le 2$ whenever the seed amplitude satisfies $\mathcal{V}(|\alpha|) = \eta_{\mathrm{mode}}|\alpha|^2/(1+|\alpha|^2) > 1/\sqrt{2}$ (i.e., $|\alpha| > 1.56$ for $\eta_{\mathrm{mode}}=1$, and $\mathcal{S}_{\mathrm{CHSH}} \approx 2.80$ at our experimental operating point $\mathcal{V}\approx 0.99$).
\end{enumerate}

\subsection{Hybrid coupling to hardware-native TF-GKP grid states for temporal error correction}
\label{sec:S7_3_TF_GKP}

Finally, a unique feature of our ENBS realization is that the discrete two-level time qubit $\mathcal{H}_T$ is natively embedded within the continuous-variable chronocyclic phase space of the $N\approx 360$-tooth SPFC supermode (Section~\ref{sec:S4_TF_vs_relational}), providing a direct bridge to bosonic time--frequency grid encodings and temporal quantum error correction.

Following the hardware-native TF-GKP framework developed in Ref.~\cite{Yoon2026a} (building on the modal time--frequency constructions of Fabre, Descamps, Milman, and co-workers), we introduce the dimensionless canonical arrival-time and frequency operators on the SPFC supermode,
\begin{equation}
	\hat{\tau} = \frac{\hat{T}}{T_{\mathrm{rep}}},
	\qquad
	\hat{\Omega} = T_{\mathrm{rep}}\hat{\omega},
	\qquad
	[\hat{\tau}, \hat{\Omega}] = i,
	\label{eq:S7_dimensionless_TF}
\end{equation}
where $T_{\mathrm{rep}} = 1/f_{\mathrm{rep}} = 4$~ns is the stabilized comb repetition period. Calibrated temporal delays $\Delta t = \tau T_{\mathrm{rep}}$ (implemented via EOM phase ramps or optical delays) and frequency translations $\Delta\omega = \Omega/T_{\mathrm{rep}}$ (implemented via AOM frequency shifts) generate the dimensionless Weyl--Heisenberg displacement operator
\begin{equation}
	\hat{D}_{\mathrm{TF}}(\tau,\Omega) = \exp\!\big[i(\Omega\hat{\tau} - \tau\hat{\Omega})\big],
	\label{eq:S7_Weyl_TF}
\end{equation}
which defines the commuting square-lattice GKP stabilizers $\hat{S}_\tau = \hat{D}_{\mathrm{TF}}(0,2\sqrt{\pi})$ and $\hat{S}_\Omega = \hat{D}_{\mathrm{TF}}(2\sqrt{\pi},0)$ and logical Pauli operators $\bar{Z}_{\mathrm{GKP}} = \hat{D}_{\mathrm{TF}}(\sqrt{\pi},0)$ and $\bar{X}_{\mathrm{GKP}} = \hat{D}_{\mathrm{TF}}(0,\sqrt{\pi})$.

When symmetric opposite EOM/AOM modulations $\pm(\delta\tau, \delta\Omega)$ are applied to the two arms of the ENBS time-qubit interferometer, the effective branch interaction $\hat{V}$ on the chronocyclic envelope becomes $\hat{V}_{\mathrm{TF}} = (\hbar/t)(\delta\Omega\hat{\tau} - \delta\tau\hat{\Omega})$, so that the composite time-qubit evolution $\hat{U}_{\mathrm{tot}} = \exp(-i\hat{H}_{\mathrm{tot}}t/\hbar)$ implements a \emph{controlled time--frequency displacement gate} between the discrete time qubit $\mathcal{H}_T$ and the continuous SPFC grid mode:
\begin{equation}
	\hat{C}_D(\delta\tau,\delta\Omega)
	= \exp\!\big[-i\,\tau_z\otimes(\delta\Omega\hat{\tau} - \delta\tau\hat{\Omega})\big]
	= \ket{+}\!\bra{+}_T\otimes\hat{D}_{\mathrm{TF}}(+\delta\tau,+\delta\Omega)
	+ \ket{-}\!\bra{-}_T\otimes\hat{D}_{\mathrm{TF}}(-\delta\tau,-\delta\Omega).
	\label{eq:S7_controlled_displacement}
\end{equation}
Setting $(\delta\tau,\delta\Omega) = (0,\sqrt{\pi})$ or $(\sqrt{\pi},0)$ and measuring the discrete time qubit in the $\tau_x/\tau_y$ basis at the output beam splitter [Eq.~\eqref{eq:S4_Wigner_time_qubit}] directly reads out the real and imaginary parts of the GKP stabilizer expectation values $\expval*{\hat{S}_\tau}$ and $\expval*{\hat{S}_\Omega}$ via first-order single-photon counting rates. Thus, the discrete ENBS time qubit developed in this work functions natively as the ancilla control subsystem required for stabilizer syndrome extraction and active temporal quantum error correction of bosonic TF-GKP logical qubits.

\end{document}